\documentclass[6pt, preprint2]{aastex}
\usepackage[]{natbib}
\begin{document}
\title{Further Evidence for Quantized Intrinsic Redshifts in Galaxies: Is the Great Attractor a Myth?}
\author{M.B. Bell\altaffilmark{1} and S.P. Comeau\altaffilmark{1}}
\altaffiltext{1}{Herzberg Institute of Astrophysics,
National Research Council of Canada, 100 Sussex Drive, Ottawa,
ON, Canada K1A 0R6}

\begin{abstract}

Evidence was presented recently suggesting that the Fundamental Plane (FP) clusters studied in the Hubble Key Project may contain quantized intrinsic redshift components that are related to those reported by Tifft. Here we report the results of a similar analysis using 55 spiral (Sc and Sb) galaxies, and 36 Type Ia supernovae (SnIa) galaxies. We find that even when many more objects are included in the sample there is still clear evidence that the same quantized intrinsic redshifts are present and superimposed on the Hubble flow. We find Hubble constants of H$_{\rm o}$ = 60.0 and 57.5 km s$^{-1}$ Mpc$^{-1}$ for the Sc and Sb galaxies respectively. For the SnIa galaxies we find H$_{\rm o}$ = 58. These values are considerably lower than the value of H$_{\rm o}$=72 reported by the Hubble Key Project, but are good in agreement with the value H$_{\rm o}$ = 60 found for intermediate redshifts using the Sunyaev-Zel'dovich (SZ) effect. Evidence is also presented that suggests that the presence of unaccounted for intrinsic redshifts may have led us incorrectly to the conclusion that a "great attractor" is needed to explain the velocity data. The 91 galaxies examined here also offer new, independent confirmations of the importance of the redshift increment z$_{f}$ = 0.62.
%These results are seen as another independent confirmation of the intrinsic %redshift model presented previously, and of the quantized "velocities" found by %Tifft in galaxies. 

\end{abstract}

\keywords{galaxies: Cosmolgy: distance scale -- galaxies: Distances and redshifts - galaxies: quasars: general}

\section{Introduction}

Recently evidence was presented \citep{bel03a} that indicated that the FP clusters studied in the Hubble Key Project \citep{fre01} may contain quantized intrinsic redshift components that are harmonically related to those reported by \citet{tif96,tif97}, and by \citet{bel02d}. When these were identified and taken into account the Hubble constant for the FP clusters was found to be H$_{\rm o}$ = 71.4 km s$^{-1}$ Mpc$^{-1}$, reduced significantly from the value of H$_{\rm o}$ = 82 reported by \citet{fre01}. This study also found that the dispersion in cluster velocities fell to near 56 km s$^{-1}$ when the intrinsic components were remove and implied that the relative uncertainty in the FP cluster distances was of the order of 1 Mpc. The main conclusion of that study was that if the radial velocities of galaxies contain an intrinsic redshift component, unless it can be identified and removed the Hubble constant obtained would be too high.

Here we describe a similar analysis carried out on three different source samples containing 91 galaxies. They include 23 Sc I, Sc I.2, Sc I.3 (hereafter Sc) galaxies, 32 Sab, Sb, Sb I-II (hereafter Sb) galaxies from \citet[Table 6 and 7]{rus02a}, and 36 SnIa galaxies from \citet[Table 6]{fre01}. For the Sc galaxies the relative uncertainty in the distance modulus is reported to be 0.09 mag \citep{rus02b}, which leads to a scatter $\sim$1 Mpc. This would allow at least the larger intrinsic redshift structure to be seen. Unfortunately the uncertainty for the Sb galaxies is larger ($\sim0.25$ mag) \citep{rus02b}, and this larger uncertainty is expected to make any intrinsic redshift quantization less obvious.

\section{Intrinsic Redshifts Defined in Previous Work}

It has been demonstrated previously from a study of the QSOs near NGC 1068, in which an attempt was made to remove all Doppler-related redshifts \citep{bel02a,bel02b,bel02c}, that the remaining intrinsic redshift components in those objects are harmonically related to the redshift z$_{f}$ = 0.62 \citep{bel03a}. It was further suggested \citep{bel02c} from clumps in the redshift distribution of presumably nearby quasars found in early surveys, that $all$ QSOs may have an intrinsic redshift component that is harmonically related of this redshift increment. These have been defined previously \citep{bel02d,bel03a} by the relation:

\begin{equation}
z_{iQ} = z_{f}[N - 0.1M_{N}]
\end{equation}
where z$_{f}$ = 0.62 is a fundamental redshift increment, $N$ is an integer, and $M_{N}$ can have only certain well-defined values defined by a quantum number $n$.

It has also been pointed out \citep{bel02d} that the most common discrete redshift components in normal galaxies, found by \citet{tif96,tif97}, are all harmonically related to z$_{f}$ = 0.62 and are defined by the relation:

\begin{equation}
z_{iG}[N,m] = \left(\frac{z_{iQ}\left[N,n_{max}\right]}{2^{m}}\right)_{m=0,1,2,3..\infty}
\end{equation}
where the quantum number $m$ represents the number of doublings (halvings) below the relevant quasar minimum intrinsic redshift component z$_{iQ}$[$N,n_{max}$] (see \citet{bel03a} for a more complete description of the intrinsic components). Values for the larger intrinsic redshift components in galaxies, and their velocity equivalents, are listed in Table 4 for $N$ = 1 to 6. 
%For each value of $N$ the relevant $n_{\rm max}$ value is given by the value in %Table 4 corresponding to $m$ = 0.

Mass density fluctuations in the Universe can cause peculiar velocities that are not related to the Hubble flow. In fact all peculiar velocities in galaxies in the near Universe are likely to be density induced because all primordial turbulence is expected to have been damped out by adiabatic expansion \citep{kra86}. Thus, if the locally produced peculiar velocities can be completely accounted for, and distances are accurate, it is argued here that any large excess redshift components remaining should be considered as possible intrinsic redshift candidates.

\section{Data Analysis}

The analysis used here follows that used previously \citep{bel03a} where a minimum is sought in the RMS deviation in source velocities, calculated relative to the nearest intrinsic redshift grid line, as the Hubble constant is varied. The intrinsic redshifts are assumed to be superimposed on the Hubble flow. The slope of the grid lines is determined by H$_{\rm o}$ and their intrinsic redshift values are given by eqn 2, and are listed here in Table 4. Eqn 2 is a function of z$_{iQ}$[$N,n_{max}$] which in turn is a function of z$_{f}$, as can be seen from eqn 1. Note that $n_{max}$ is defined as the largest value $n$ can have before a quasar intrinsic redshift becomes a blueshift.

%If it is assumed a) that galactic distances are accurately known, b) that all %peculiar velocities due to local density fluctuations can be accurately %accounted for, and c) that the observed redshifts are entirely Doppler-related %and due to the Hubble flow, (i.e. no intrinsic components present) then, if %H$_{\rm o}$ = 72, galaxies should all fall along the solid line in the Hubble %plot in Fig. 1. If there are intrinsic components present in some galaxies with %the discrete values given by equation ($N$ = 0 for galaxies), then these %sources will fall on the appropriate dashed lines in Fig. 1.

\subsection{Sc I, Sc I.2, Sc I.3 Galaxies}

\begin{figure}
\hspace{-1.0cm}
\vspace{-2.0cm}
\epsscale{1.0}
\plotone{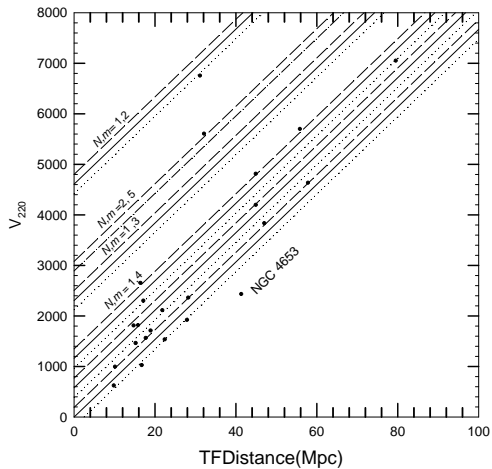}
\caption{\scriptsize{V$_{220}$ plotted vs Tulley-Fisher distance for Sc galaxies in Table 6 of \citet{rus02a}. Solid lines show Hubble lines for sources with discrete velocities of 0, 580.1, 1158.0, 2314.4, and 4610.8 km s$^{-1}$, given by equation B1 of \citet{bel03a}. Dashed and dotted lines show Hubble plots for each of the above intrinsic components shifted by $\pm190$ km s$^{-1}$. All grid lines assume H$_{\rm o}$ = 76.7. \label{fig1}}}
\end{figure}

\begin{deluxetable}{ccccc}
\tabletypesize{\scriptsize}
\tablecaption{Spiral Galaxies\tablenotemark{a}. \label{tbl-1}}
\tablewidth{0pt}
\tablehead{
\colhead{Source} & \colhead{R.A.} & \colhead{Dec.} & \colhead{Red} & \colhead{Blue} 
}
\startdata
NGC 309 & 0056  & -0954  & x  & \\
NGC 958 & 0230  & -0256  & x  &   \\
NGC 1232 & 0310  & -2034 & x  &   \\
NGC 1376 & 0337 & -0502  & x  &  \\
NGC 2207 & 0616  & -2122 & x  &   \\
NGC 2280 & 0644  & -2738 & x  &   \\
NGC 2835 & 0917 & -2221 &    & x   \\
NGC 2942 & 0939 & 3400  & x  &  \\
NGC 2955 & 0941 & 3552  &   & x   \\
NGC 2997 & 0944 & -3111 & x  &   \\
NGC 2998 & 0948 & 4404  & x  &   \\
NGC 3294 & 1036 & 3719 &    & x   \\
NGC 3464 & 1054 & -2104  & x  &   \\
NGC 3614 & 1118 & 4544 & x  &   \\
NGC 3893 & 1148 & 4842 &   & x   \\
NGC 4254 & 1218 & 1424 &   & x  \\
NGC 4321 & 1222 & 1549  &  & x   \\
NGC 4535 & 1234 & 0811 & zero & zero   \\
NGC 4653 & 1243 & -0033 &  & x   \\
NGC 5161 & 1329 & -3310 &   & x  \\
NGC 5230 & 1335 & 1340  &   & x  \\
NGC 5364 & 1356 & 0500  &  & x   \\
NGC 6118 & 1621 & -0217  &   & x   \\
\enddata 
\tablenotetext{a}{from Table 6 of \citet{rus02a} for H$_{\rm o}$ = 76.7}

\end{deluxetable}

In Fig. 1 the V$_{220}$ velocities (corrected for Virgo infall) for the 23 Sc galaxies listed in Table 6 of \citet{rus02a} are plotted vs their Tulley-Fisher distance. Solid lines show where galaxies with discrete velocities of 0, 580, 1158, 2344, and 4611, km s$^{-1}$ ($N$ = 1 state) would lie if H$_{\rm o}$ = 76.7. The dot-dash line represents the location of the 2890 km s$^{-1}$ velocity of the [2,5] line in the $N$ = 2 state. On each side of these lines, the dashed (redshifted) and dotted (blueshifted) lines show these discrete redshift lines shifted by $\pm190$ km s$^{-1}$ respectively. One source (NGC 4653) cannot be fitted to any line with a Hubble slope of 76.7 km s$^{-1}$ Mpc$^{-1}$. Of the remaining 22 sources, 20 fall very close to, or on top of, one of the split z$_{\rm iG}$[1,m] gridlines. One source falls on the red split z$_{\rm iG}$[2,5] gridline. Only one of the 23 galaxies is located on a solid line. Thus these sources which have been corrected only for Virgo infall are clearly still split about the Hubble flow, represented by the solid intrinsic redshift gridlines, with a velocity splitting of $\pm190$ km s$^{-1}$. Assuming that this is a real velocity splitting caused by a directed motion, it should be possible to find the direction of motion by examining the locations of the sources.

A close look at the co-ordinates of the galaxies reveals that the blueshifted ones lie mainly in the Virgo (12$^{\rm h}$) hemisphere while most of those on the redshifted (dashed) lines lie in the anti-Virgo (0$^{\rm h}$) hemisphere. This can be seen clearly in Table 1 where the galaxies and their co-ordinates are listed. There is some ambiguity for the first dashed and dotted lines above 0 because they lie close together but Fig 1, combined with Table 1, shows what appears to be motion of our Galaxy, relative to the other sources, towards the hemisphere of the sky containing Virgo. 

In Fig 2 the Right Ascension of each source is plotted vs its Declination. Redshifted sources are indicated by filled circles and blueshifted ones by open circles. It is apparent from this plot that the crossover R.A. is close to 10$^{\rm h}$ suggesting that the directed motion is towards R.A. = 16$^{\rm h}$. In Table 1, over 80 percent of the redshifted sources are in the R.A. = 4$^{\rm h}$ hemisphere and over 80 percent of the blueshifted sources are in the 16$^{\rm h}$ hemisphere. This direction is within about 30\arcdeg of the direction of flow found for the Great Attractor (GA) in the model of \citet{lyn88} where the motion was found to be towards $l$ = 309\arcdeg; $b$ = +18\arcdeg. It should be clearly understood, however, that the motion suggested by Fig 1 and Table 1 appears to be one involving the Galaxy only, unlike the general flow envisioned in the GA model.

It is important to note that these data show that there are two independent effects involved. These are a), a motion of our Galaxy relative to the Hubble flow, directed towards a region of sky that has previously been assumed to contain an over-density of matter (the Great Attractor) and b), an excess intrinsic redshift component superimposed on the Hubble flow. The latter of these has previously been interpreted as an excess of positive peculiar velocities relative to the Hubble flow, in the same (Great Attractor) direction. The first of these appears as a blueshift of sources in the GA direction and a redshift in the anti-GA direction, \em relative to the Sc galaxies \em. The second appears as an excess redshift in all directions. These, together with the Virgo infall velocity, would have previously been combined and confused. This is discussed in more detail in section 9 below.

\begin{figure}
\hspace{-1.0cm}
\vspace{-1.0cm}
\epsscale{1.0}
\plotone{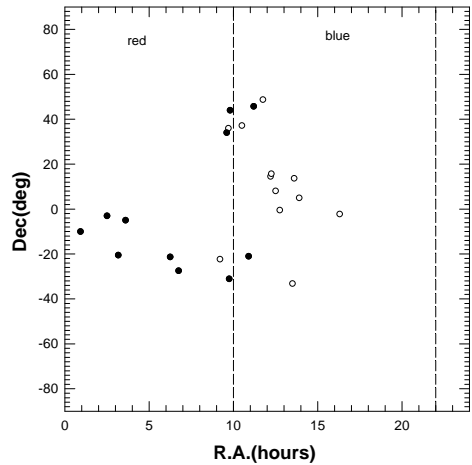}
\caption{\scriptsize{Right Ascension of Sc galaxies from Table 6 of \citet{rus02a} plotted vs declination. Filled circles are for redshifted galaxies (dashed lines in Fig 1) and open circles are for blueshifted ones (dotted lines in Fig 1). \label{fig2}}}
\end{figure}

\begin{figure}
\hspace{-1.0cm}
\vspace{-1.0cm}
\epsscale{1.0}
\plotone{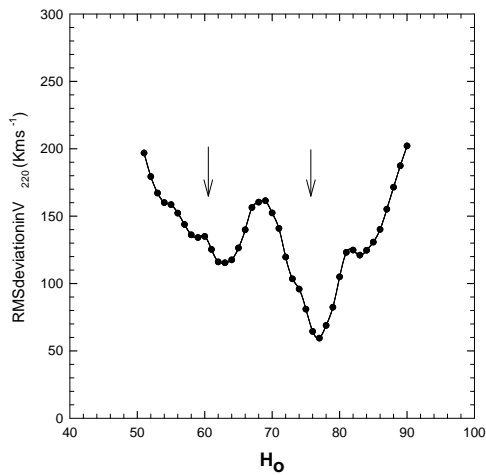}
\caption{\scriptsize{RMS deviation in V$_{220}$ from split grid lines in Fig 1 as a function of H$_{\rm o}$ for the z$_{f}$ = 0.62.  \label{fig3}}}
\end{figure}

\begin{figure}
\hspace{-1.0cm}
\vspace{-2.0cm}
\epsscale{1.0}
\plotone{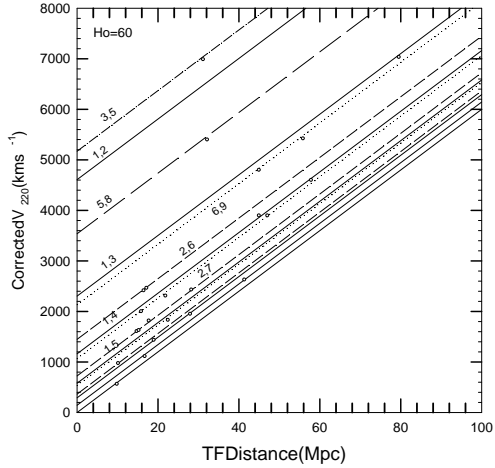}
\caption{\scriptsize{Corrected velocities plotted vs Tulley-Fisher distance for Sc galaxies in Table 6 of \citet{rus02a}. Hubble lines are shown for intrinsic redshifts corresponding to $N,m$ = 1,2 1,3 1,4 1,5 1,6 1,7 2,6 2,7 2,8 3,5 5,8 6,9 6,10 and 6,11 listed in Table 4.  \label{fig4}}}
\end{figure}

\begin{figure}
\hspace{-1.0cm}
\vspace{-1.0cm}
\epsscale{1.0}
\plotone{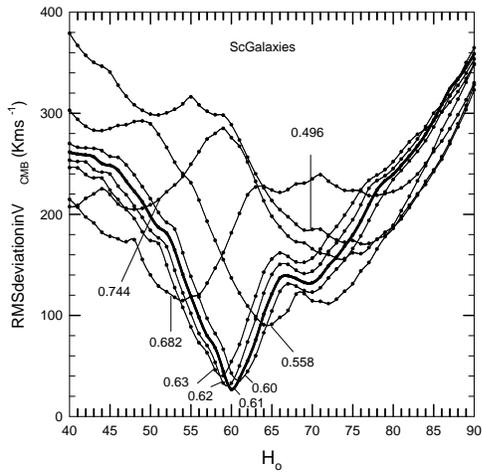}
\caption{\scriptsize{RMS deviation in V$_{\rm 220}$ velocities corrected for 300 km s$^{-1}$ motion in the R.A = 16$^{\rm h}$; DEC = 0$\arcdeg$ direction from grid lines in Fig 1 as a function of H$_{\rm o}$ for the z$_{f}$ values indicated.  \label{fig5}}}
\end{figure}

\begin{figure}
\hspace{-1.0cm}
\vspace{-1.0cm}
\epsscale{1.0}
\plotone{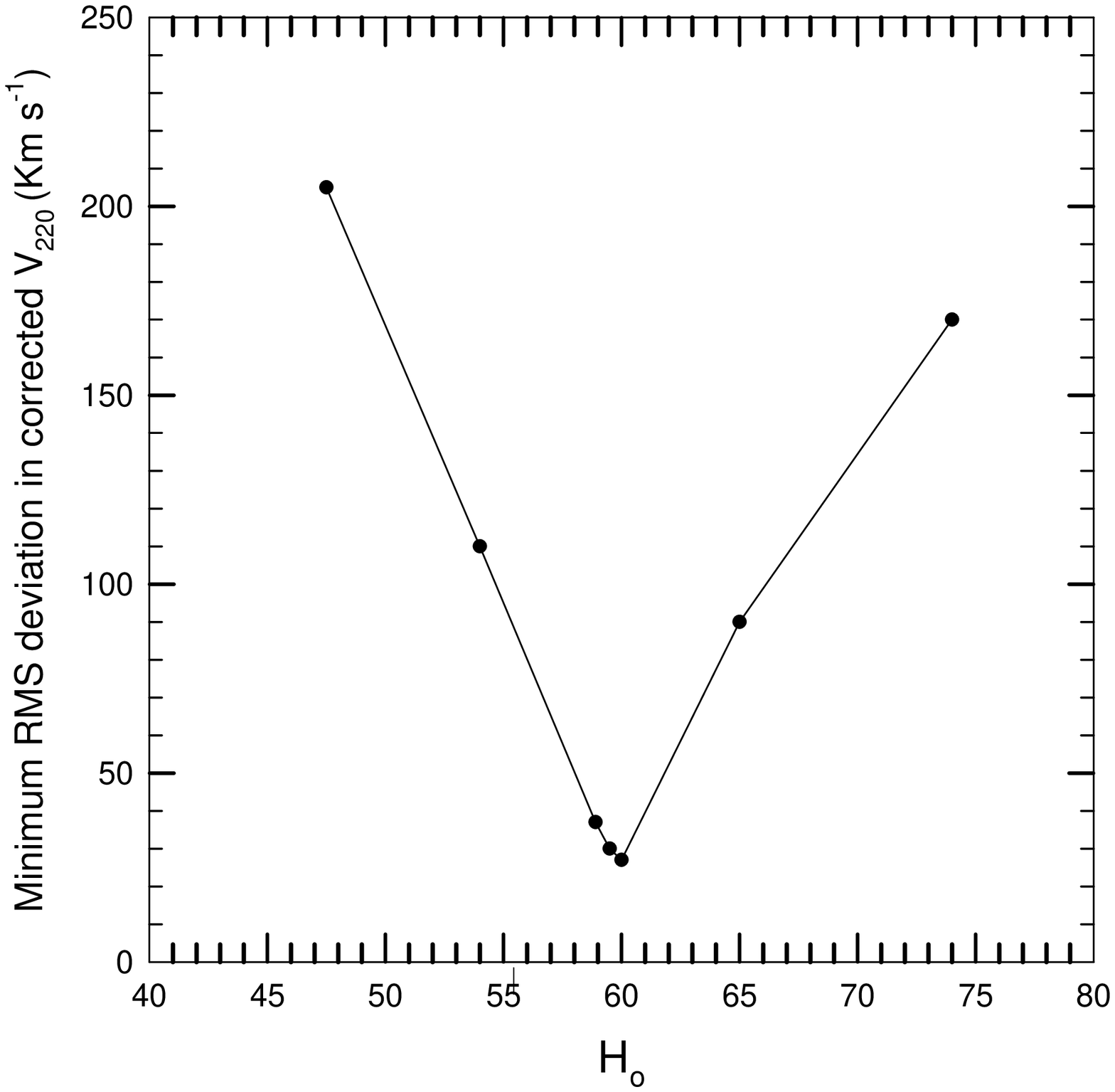}
\caption{\scriptsize{Plot of minimum RMS deviation in corrected V$_{220}$ velocities vs Ho for data from Table 6 of \citet{rus02a}.  \label{fig6}}}
\end{figure}

\begin{figure}
\hspace{-1.0cm}
\vspace{-1.0cm}
\epsscale{1.0}
\plotone{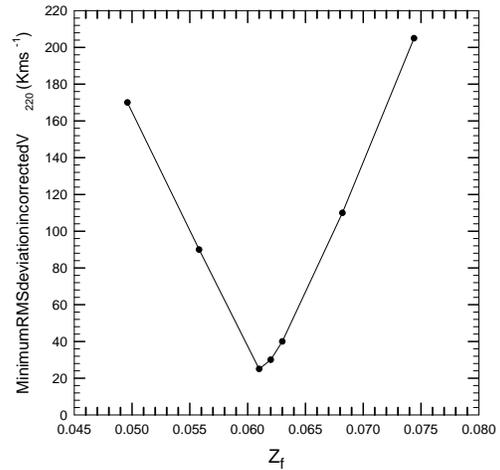}
\caption{\scriptsize{Plot of minimum RMS deviation in corrected V$_{220}$ velocities vs z$_{f}$ for data from Table 6 of \citet{rus02a}. \label{fig7}}}
\end{figure}

As was done in earlier work, we next calculated the RMS deviation in velocities relative to the grid lines (this time using the V$_{220}$ velocities and split grid lines) for a range of Hubble constants between 50 and 90 and z$_{f}$ = 0.62. The result is shown in Fig 3 where the arrows indicate two minimums in the RMS deviation - one near H$_{\rm o}$ = 60 and the other near H$_{\rm o}$ = 77. 

If, as suggested above, our Galaxy is moving toward a region in the sky near R.A. = 16$^{\rm h}$ with a velocity V$_{\rm MW}$, it is possible, using the source co-ordinates, to correct each source's measured velocity for this motion. It is apparent from Fig. 2 that approximately half the sources lie at angular distances greater than 45$\arcdeg$ from the direction of motion. Thus it can be predicted that the velocity of our Galaxy in the 16$^{\rm h}$ direction required to produce the splitting visible in Fig 1 will be considerably in excess of V$_{\rm MW}$ = 200 km s$^{-1}$.

\begin{figure}
\hspace{-1.0cm}
\vspace{-1.0cm}
\epsscale{1.0}
\plotone{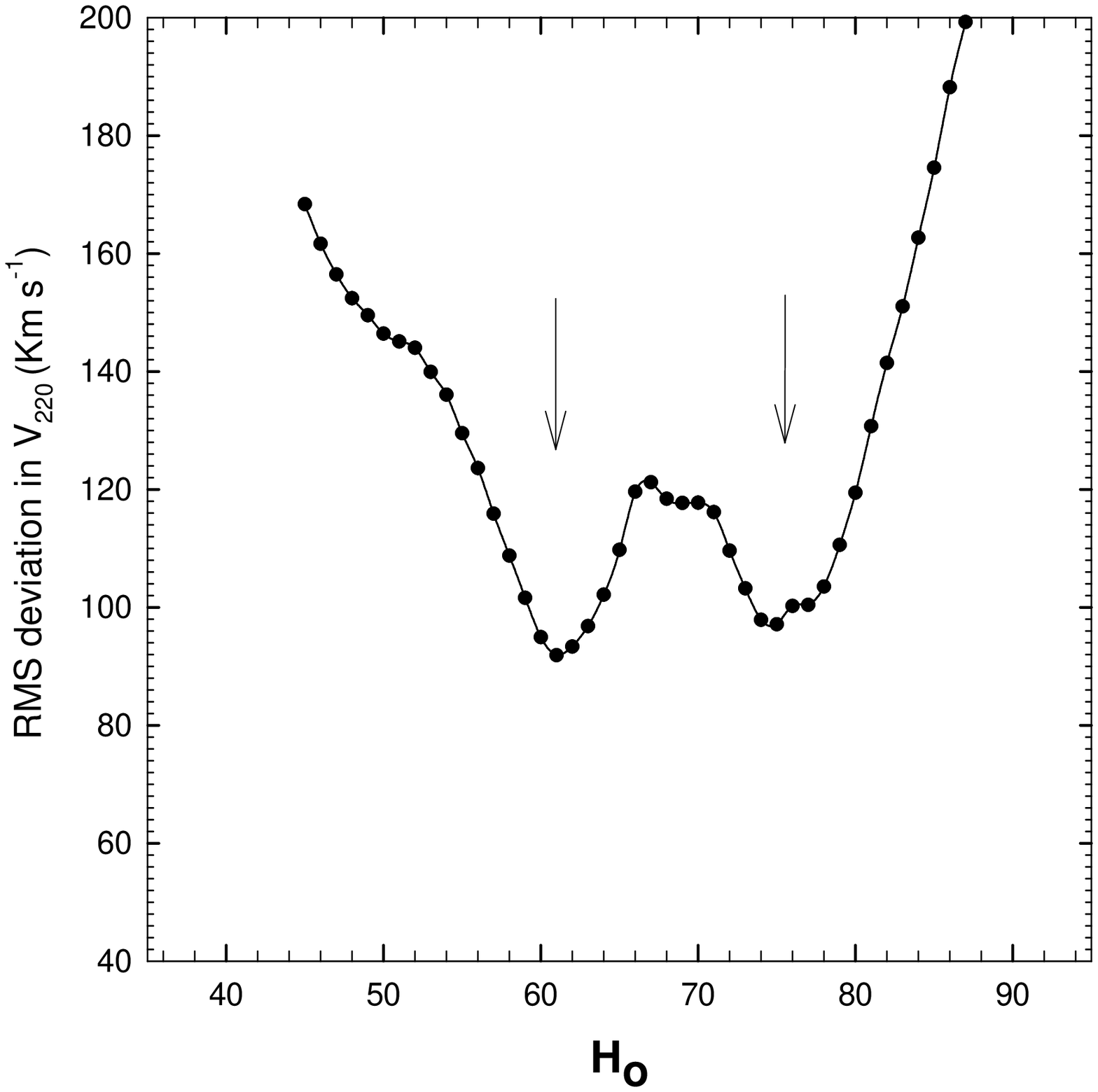}
\caption{\scriptsize{Plot of RMS deviation in V$_{\rm 220}$ velocities about the split grid lines as a function of the Hubble constant for z$_{f}$ = 0.62. Data are from Table 7 of \citet{rus02a}. \label{fig8}}}
\end{figure}

\begin{figure}
\hspace{-1.0cm}
\vspace{-1.0cm}
\epsscale{1.0}
\plotone{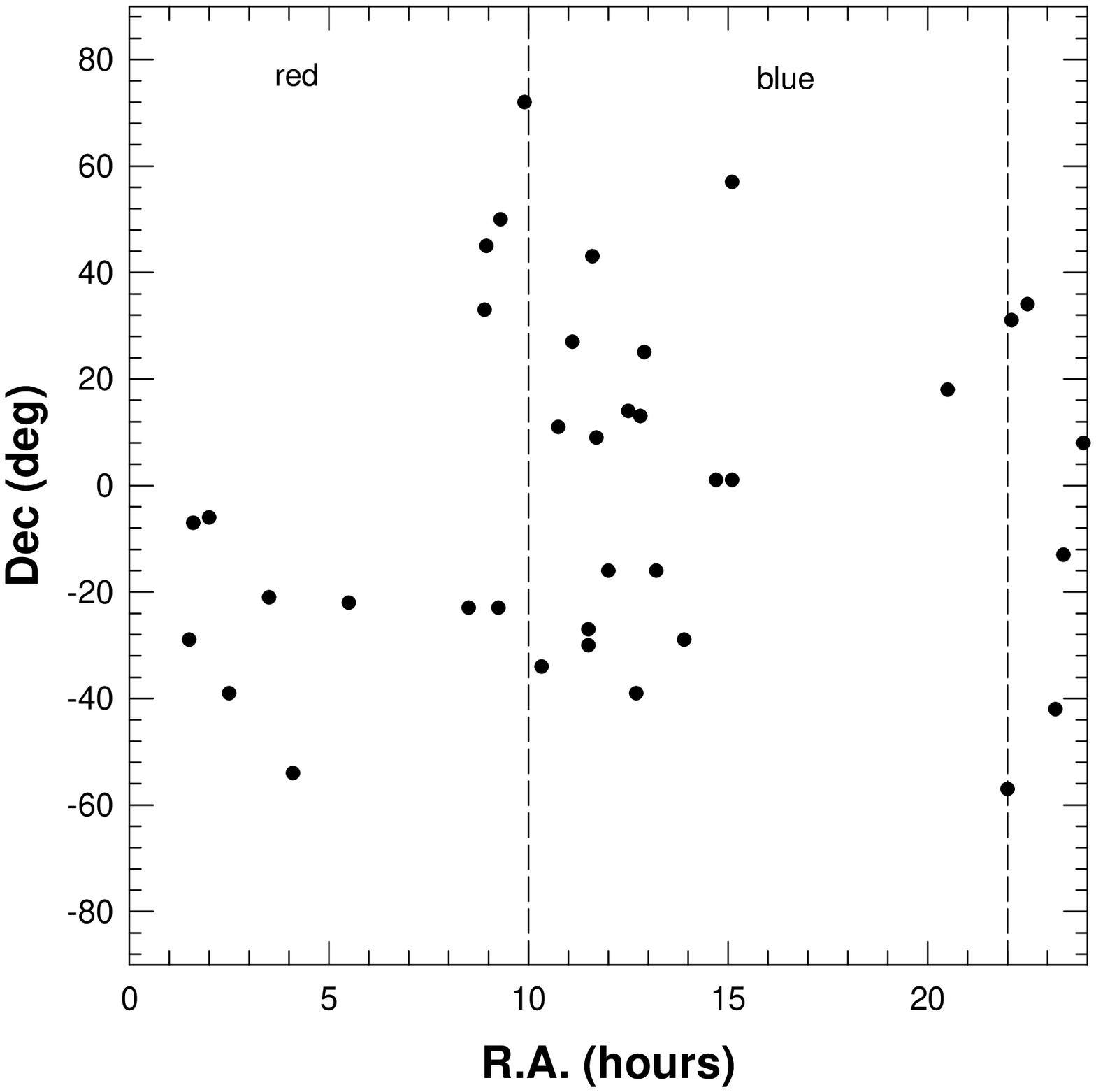}
\caption{\scriptsize{Plot of DEC vs R.A. for Sb galaxies in Table 7 of \citet{rus02a}. Dashed lines separate redshifted and blueshifted hemispheres as defined by the assumed direction of motion of the Galaxy. \label{fig9}}}
\end{figure}

In Fig 4 the source velocities have again been plotted vs distance as in Fig 1, after correcting for the motion of our Galaxy towards a point in the sky at R.A. = 16$^{\rm h}$; DEC = $0\arcdeg$. In making the corrections, several different V$_{\rm MW}$ velocities were tried between 200 and 400 km s$^{-1}$ and a velocity V$_{\rm MW}$ = 300 km s$^{-1}$ was determined to be the optimum value by producing the best fit to the grid lines. In Fig 4, the slope of the grid lines represents a Hubble constant of 60. NGC 4653, which we were unable to fit to a grid line in Fig 1 is now located on the 1,7 line. There are several rows of sources that exhibit a slope close to 60 and they can all be fitted to one of the predicted gridlines. At least seven of the 23 sources appear to be N = 1 sources, 7 are N = 2 sources, 1 is an N = 3 source, 2 are N = 5 sources and 6 are N = 6 sources. Note that if a small increase in the distance uncertainty is assumed, the 6, N = 6 sources could equally as well be counted as N = 1 sources.  
In this case 20 of the 23 sources would be in either the $N$ = 1 or 2 states.
 
We then calculated the RMS deviation of the corrected velocities from the grid lines in Fig. 4 for a range of H$_{\rm o}$ values and the curves are shown in Fig 5 for several different z$_{f}$ values. The heavy dark curve represents the value z$_{f}$ = 0.61. It is now apparent that the RMS dip near H$_{\rm o}$ = 60 in Fig. 3 has deepened significantly after correcting for the motion assumed above, while the dip previously seen near H$_{\rm o}$ = 77 has disappeared. We also looked at cases where the direction of motion was assumed to be towards 14$^{\rm h}$ and 18$^{\rm h}$. In both cases the fit deteriorated slightly.

In Fig 6 the minimum RMS deviation in each of the z$_{f}$ curves is plotted as a function of H$_{\rm o}$ and shows that the lowest minimum, or best fit, occurs for H$_{\rm o}$ = 60.0.
In Fig 7 the minimum RMS deviation in each curve is plotted as a function of z$_{f}$ and shows that the best fit occurs for z$_{f}$ = 0.613.

\begin{figure}
\hspace{-1.0cm}
\vspace{-2.0cm}
\epsscale{1.0}
\plotone{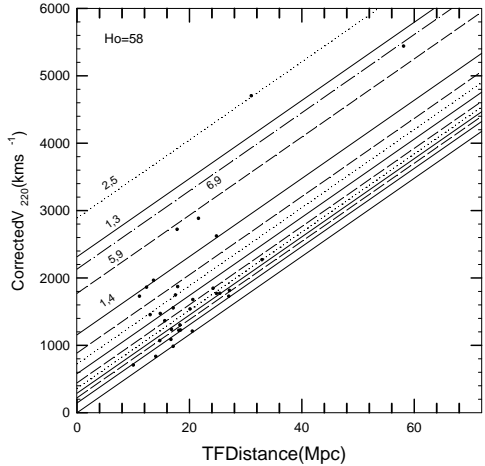}
\caption{\scriptsize{Corrected velocities plotted vs Tulley-Fisher distance for Sb galaxies in Table 7 of \citet{rus02a}. Solid lines show Hubble lines for sources with discrete velocities of 0, 145.1, 290.2, 580.1, 1158.0, and 2314.4, (N = 1 sources); 2890.2, and 725.2 (N = 2 sources); 1772, 888, and 444 (N = 5 sources); 2133 (N = 6 sources) km s$^{-1}$, as listed in Table 4. \label{fig10}}}
\end{figure}

\begin{figure}
\hspace{-1.0cm}
\vspace{-1.0cm}
\epsscale{1.0}
\plotone{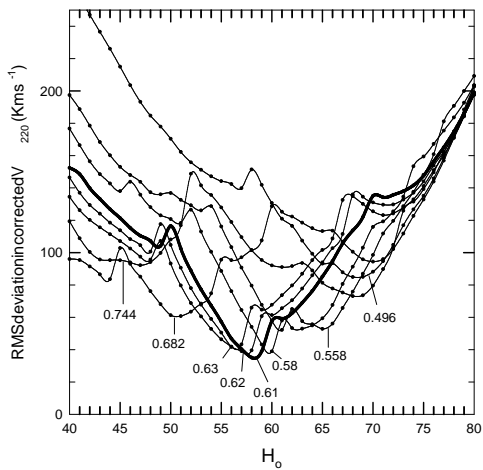}
\caption{\scriptsize{Plot of RMS deviation in corrected V$_{\rm 220}$ velocities as a function of the Hubble constant for several different z$_{f}$ values. Data are for Sb galaxies in Table 7 of \citet{rus02a}. \label{fig11}}}
\end{figure}

\subsection{Sab, Sb, Sb I-II, Sb II Galaxies}

In Fig 8, the RMS deviation in V$_{220}$ about the split grid lines is plotted vs H$_{\rm o}$ for the 36 Sb galaxies listed in Table 7 of \citet{rus02a}, as was done for the Sc galaxies in Fig 3. Again two RMS minimums are
visible near H$_{\rm o}$ = 60 and H$_{\rm o}$ = 77. The blue- and red-shifted sources correlated best with the Virgo and anti-Virgo hemispheres this time for the H$_{\rm o}$ value near 60, instead of 77. Although this was a mystery at first, the explanation soon became apparent. In Fig 9 the DEC for each source in Table 7 of \citet{rus02a} is plotted vs R.A. It is apparent from this figure that more than 75 percent of the sources are located at angles greater than 45$\arcdeg$ (R.A. $> 3^{\rm h}$) from the direction of motion ($16^{\rm h}$) and the corrections necessary for these sources will therefore be small, unlike the situation for the Sc galaxies. Thus, since most of the Sb galaxies will require little velocity correction, the best fit is expected to occur near the same value found above for the Sc galaxies after correction (H$_{\rm o}$ = 60) if the results are real. Since the distance accuracy is known to be better for the Sc galaxies, it also explains why, for uncorrected velocities, the RMS dip near H$_{\rm o}$ = 60 for the Sb sources in Fig. 8 is a better fit than the corresponding feature in Fig 3.

In Fig 10, the Sb galaxies have been plotted vs distance after correcting for an assumed motion of 300 km s$^{-1}$ towards R.A. = 16$^{\rm h}$; DEC = $0\arcdeg$, as was done above for the Sc galaxies.
Since the distance uncertainties are larger for the Sb galaxies it is not surprising to find that the fit to the predicted grid lines is not as good as for the Sc galaxies. Four sources fall below the zero intrinsic redshift Hubble line and this is further confirmation of the larger distance uncertainty for the Sb galaxies. These sources (NGC 1515, NGC 2683, NGC 4050, NGC 4548), were omitted from the fit and are not included in Fig 10.

In Fig 11 the RMS deviation in the corrected velocities is plotted as a function of H$_{\rm o}$ for the remaining 32 Sb galaxies, assuming V$_{\rm MW}$ = 300 km s$^{-1}$ as above. Curves for z$_{f}$ values between 0.49 and 0.74 are shown. The heavy dark curve is for z$_{f}$ = 0.61. Fig 11 shows that there is now only one RMS minimum visible as was found for the Sc data in Fig 5. 

The minimum RMS values found for the corrected velocities in Fig 11 are plotted vs H$_{\rm o}$ and z$_{f}$ in Figs 12 and 13 respectively. These curves show that for the Sb galaxies the lowest RMS value occurs at H$_{\rm o}$ = 57.5 and z$_{f}$ = 0.613.

\section{Fitting To a Complete Set of Grid Lines}

In the above analysis the intrinsic redshift grid lines used in the fit were chosen by first observing the Hubble plots (Figs 4 and 10) looking for rows of sources that might define a Hubble slope, and then choosing intrinsic redshifts from Table 4 that would fit the source distribution. Although this approach was successful, it does not rule out the possibility that other reasonable fits might be obtained if a different set of gridlines were used. To make certain that we had found the only credible fit, we redid the analysis using a more complete set of grid lines from Table 4, insuring that all lines ($m$ values) required to cover the entire range of intrinsic velocities covered by the data, were included. This is referred to hereafter as the 'universal' set of gridlines and it contained the following intrinsic redshifts: {$N$=1; $m$=2,3,4,5,6,7}, {$N$=2; $m$=4,5,6,7,8}, {$N$=3; $m$=5,6,7,8,9}, {$N$=5; $m$=8,9,10,11}, {$N$=6; $m$=8,9,10,11,12,13}. Since no lines of the $N$ = 4 state have been detected, and because they are located too close to the lines of the $N$ = 3 state to be resolved with the present distance uncertainty, the $N$ = 4 state lines were not included in the universal grid.

The universal grid is shown in Fig 14. With this many lines it seemed reasonable to assume that other, random fits might be found. In this case there are several requirements that would have to be met before a low RMS value can be considered a valid fit. For example, the profile of the fit must be narrow, with its width in H$_{\rm o}$ set approximately by the range of distances covered by the data (see Fig 5 of \citet{bel03a}). Also, it was assumed that if a good fit was found at some value of z$_{f}$ in one set of data, to be considered valid, one must also appear at the same z$_{f}$ value in the other. If it did not, it would be assumed to be a random fit. However, as will be seen below, this initial assumption turned out to be unnecessary.

In each case a range of H$_{\rm o}$ values from 40 to 80 was covered. Figs 15 and 16 show the RMS deviation curves obtained for z$_{f}$ = 0.62, for the Sc and Sb galaxies respectively. Even though many more grid lines have been used, for z$_{f}$ = 0.62, both data sets still show the best fits near Hubble constants of 60 and 57.5 for the Sc and Sb galaxies respectively. However, it is noted that there is a second reasonable fit in both data sets corresponding to a Hubble constant approximately 3 Hubble units below the best fit. The two H$_{\rm o}$ values where good fits (minimums) are seen in Fig 15 and 16 are indicated by M1 and M2.

As was done earlier, additional curves were obtained for z$_{f}$ values ranging from 0.47 to 0.74. These curves are shown in Fig 17 for the Sc galaxies. First, it is apparent from this figure that \em over the entire range of z$_{f}$ values covered, the best fit occurs for z$_{f}$ values near 0.62$\pm$ 0.01. \em The gray curves in the figure show how the \em minimum \em values obtained for the M1 and M2 features move in H$_{\rm o}$ as z$_{f}$ is varied. For the M1 feature the lowest minimum is obtained near H$_{\rm o}$ = 60, while the lowest minimum of the M2 feature occurs near H$_{\rm o}$ = 55. For the Sb galaxies the lowest minimum for the M1 feature was found to occur near 57, and that of the M2 feature near 53.

The manner in which the \em minimum \em RMS deviation in the M1 feature changes with z$_{f}$ is plotted in Fig 18 for the Sc and Sb datasets. The heavy solid line represents the Sc galaxies. Note that the values for the Sb galaxies have been scaled and normalized so both can be fitted into the same figure. From this figure it is apparent that the best fit for the M1 feature occurs at z$_{f}$ = 0.613 in both the Sc and Sb galaxy groups. Fig 19 is a similar plot for the M2 feature and shows that while the best fit for the Sc galaxies occurs near 0.634, the best fit for the Sb galaxies is near 0.652. Although the z$_{f}$ values for the M2 feature differ slightly between the Sc and Sb galaxies, the difference may not be large enough to allow us to rule out the M2 feature as a valid one when the larger distance uncertainties in the Sb group are considered. However, for the purpose of this investigation it makes no difference whether the M2 feature is included or rejected since both cases give a result that agrees well with the predicted z$_{f}$ value. 

\section{Summary of Group Parameters}

The results we obtained for the two groups of spiral galaxies studied here are summarized in Table 2. If only the M1 feature is valid, the H$_{\rm o}$ and z$_{f}$ values are in excellent agreement with those obtained earlier using only a selected set of gridlines, where H$_{\rm o}$ = 60.0 for Sc galaxies, 57.5 for Sb galaxies, and z$_{f}$ =0.613 for both.  If both features are assumed to be valid, the mean values obtained are H$_{\rm o}$ = 58.75 for Sc galaxies and 56.25 for Sb galaxies. The mean values for z$_{f}$ are 0.623 for Sc galaxies and 0.632 for Sb galaxies. Both interpretations agree with the predicted value of z$_{f}$ = 0.62 $\pm$0.01. These values also agree with the value of 0.61 found previously for the Fundamental Plane clusters. This is shown again in Fig. 20 where the minimum RMS vs z$_{f}$ curves for the M1 feature and the FP result are all included on the same plot.

It seems likely that the M2 feature is produced by the same source distribution that produces the M1 feature, since the slight decrease in H$_{\rm o}$ found for the M2 feature will be compensated for by the corresponding small increase in z$_{f}$. This double-featured fit could be due to a small error in the magnitude and/or direction of the motion of our Galaxy assumed in making the V$_{\rm MW}$ velocity corrections, but this is of no consequence to this investigation. We conclude that there is strong evidence for the presence of intrinsic redshifts in both the Sc and Sb galaxy data, and that these redshifts are identical to those found previously by us and by Tifft in galaxies.

Thus, after correcting the V$_{\rm 220}$ velocities for a V$_{\rm MW}$ = 300 km s$^{-1}$ motion of our galaxy in a direction towards R.A = 16$^{\rm h}$; DEC = $0\arcdeg$, the Hubble constants for the Sc and Sb galaxies are found to be similar, and near H$_{\rm o}$ = 58 km s$^{-1}$ Mpc$^{-1}$.

\begin{figure}
\hspace{-1.0cm}
\vspace{-1.5cm}
\epsscale{1.0}
\plotone{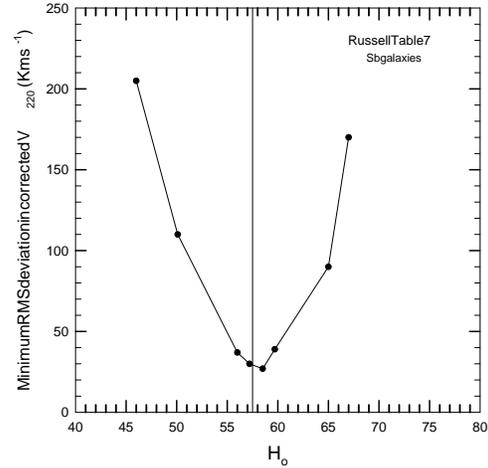}
\caption{\scriptsize{Plot of minimum RMS deviation in corrected V$_{\rm 220}$ velocities as a function of the Hubble constant for Sb galaxies in Table 7 of \citet{rus02a}. \label{fig12}}}
\end{figure}

\begin{figure}
\hspace{-1.0cm}
\vspace{-1.5cm}
\epsscale{1.0}
\plotone{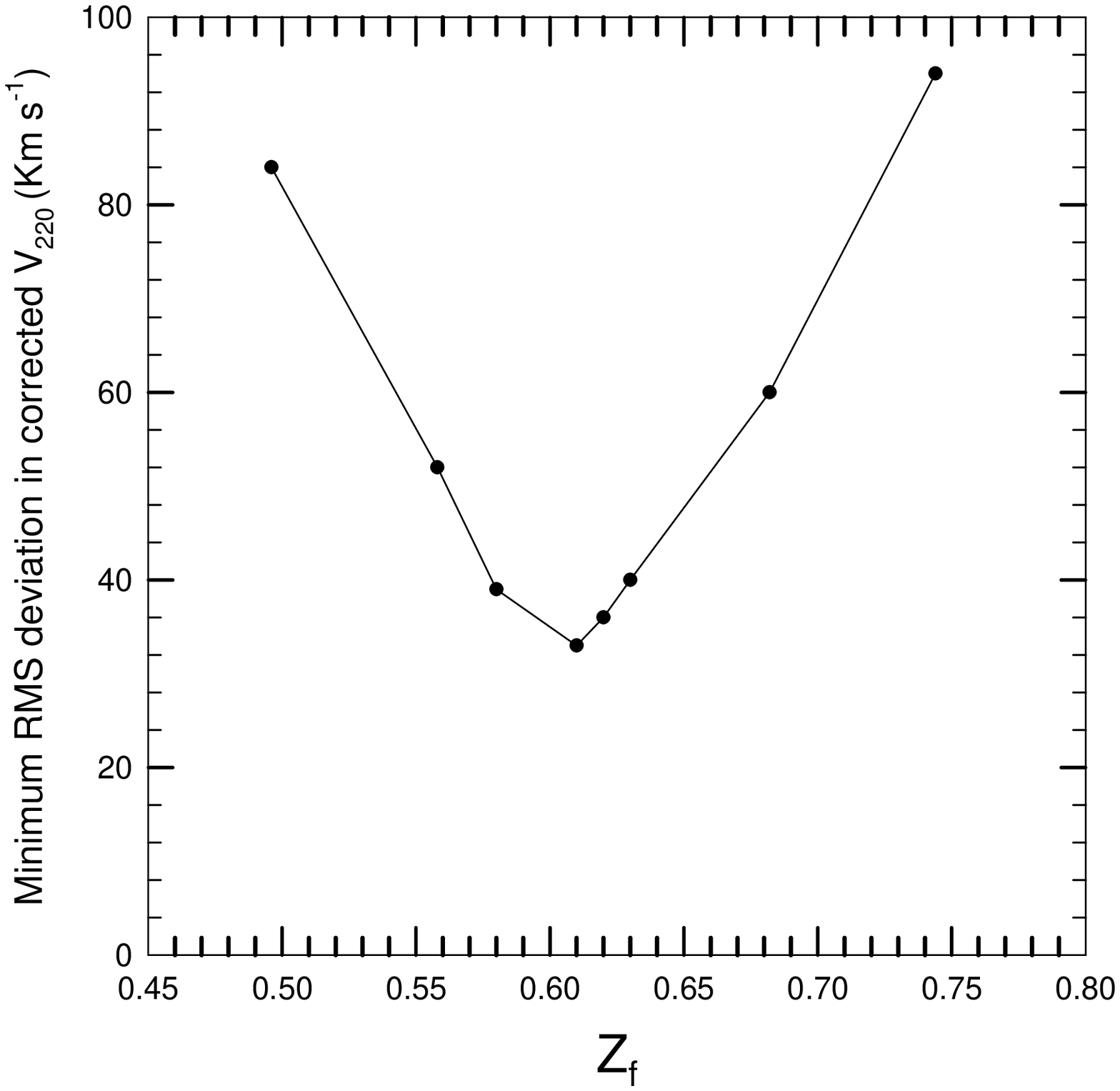}
\caption{\scriptsize{Plot of minimum RMS deviation in corrected V$_{\rm 220}$ velocities as a function of z$_{f}$ for Sb galaxies in Table 7 of \citet{rus02a}. \label{fig13}}}
\end{figure}

\begin{figure}
\hspace{-1.0cm}
\vspace{-2.0cm}
\epsscale{1.0}
\plotone{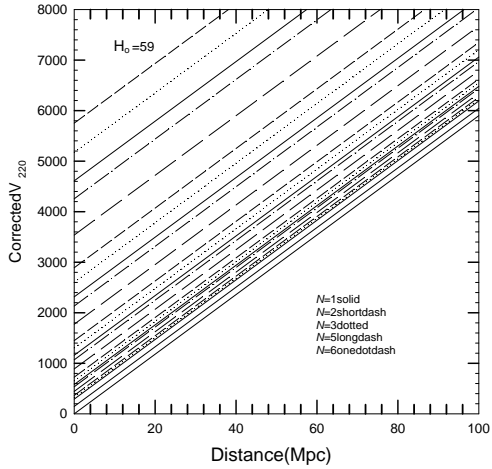}
\caption{\scriptsize{Grid lines used for the Universal grid system described in the text. \label{fig14}}}
\end{figure}

\begin{figure}
\hspace{-1.0cm}
\vspace{-1.0cm}
\epsscale{1.0}
\plotone{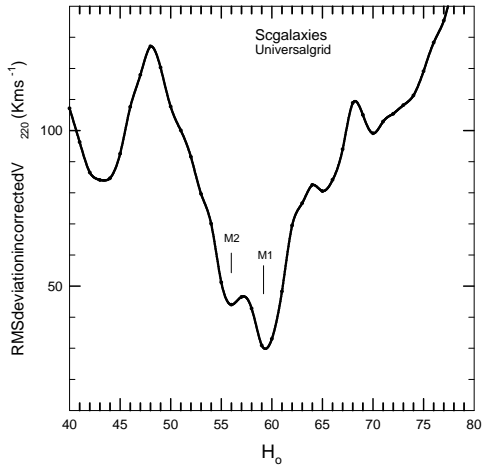}
\caption{\scriptsize{Plot of RMS deviation in corrected V$_{\rm 220}$ velocities as a function of the Hubble constant using the universal grid. Data are for Sc galaxies from Table 6 of \citet{rus02a}. The plot is for z$_{f}$ = 0.62. Features showing good fits are labeled M1 and M2. \label{fig15}}}
\end{figure}

\begin{figure}
\hspace{-1.0cm}
\vspace{-1.0cm}
\epsscale{1.0}
\plotone{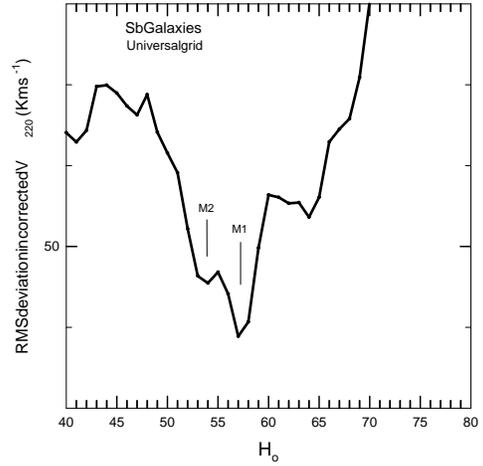}
\caption{\scriptsize{Plot of RMS deviation in corrected V$_{\rm 220}$ velocities as a function of the Hubble constant for Sb galaxies in Table 7 of \citet{rus02a}. The plot is for z$_{f}$= 0.62 \label{fig16}}}
\end{figure}

\begin{figure}
\hspace{-1.0cm}
\vspace{-1.0cm}
\epsscale{1.0}
\plotone{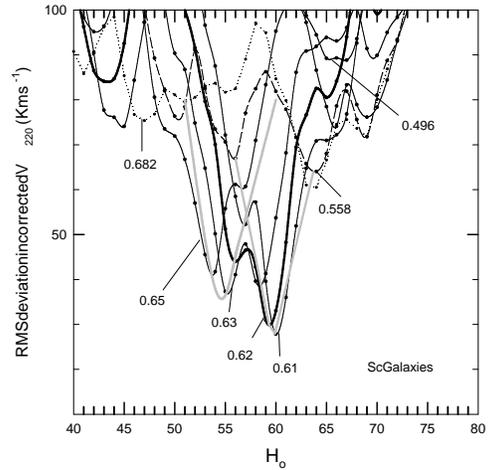}
\caption{\scriptsize{Plot of RMS deviation in corrected V$_{\rm 220}$ velocities as a function of H$_{\rm o}$ for various z$_{f}$ values as indicated. Data are for Sc galaxies in Table 6 of \citet{rus02a}. The gray curves trace the minimums of the M1 and M2 features as H$_{\rm o}$ changes. \label{fig17}}}
\end{figure}

\begin{figure}
\hspace{-1.0cm}
\vspace{-1.0cm}
\epsscale{1.0}
\plotone{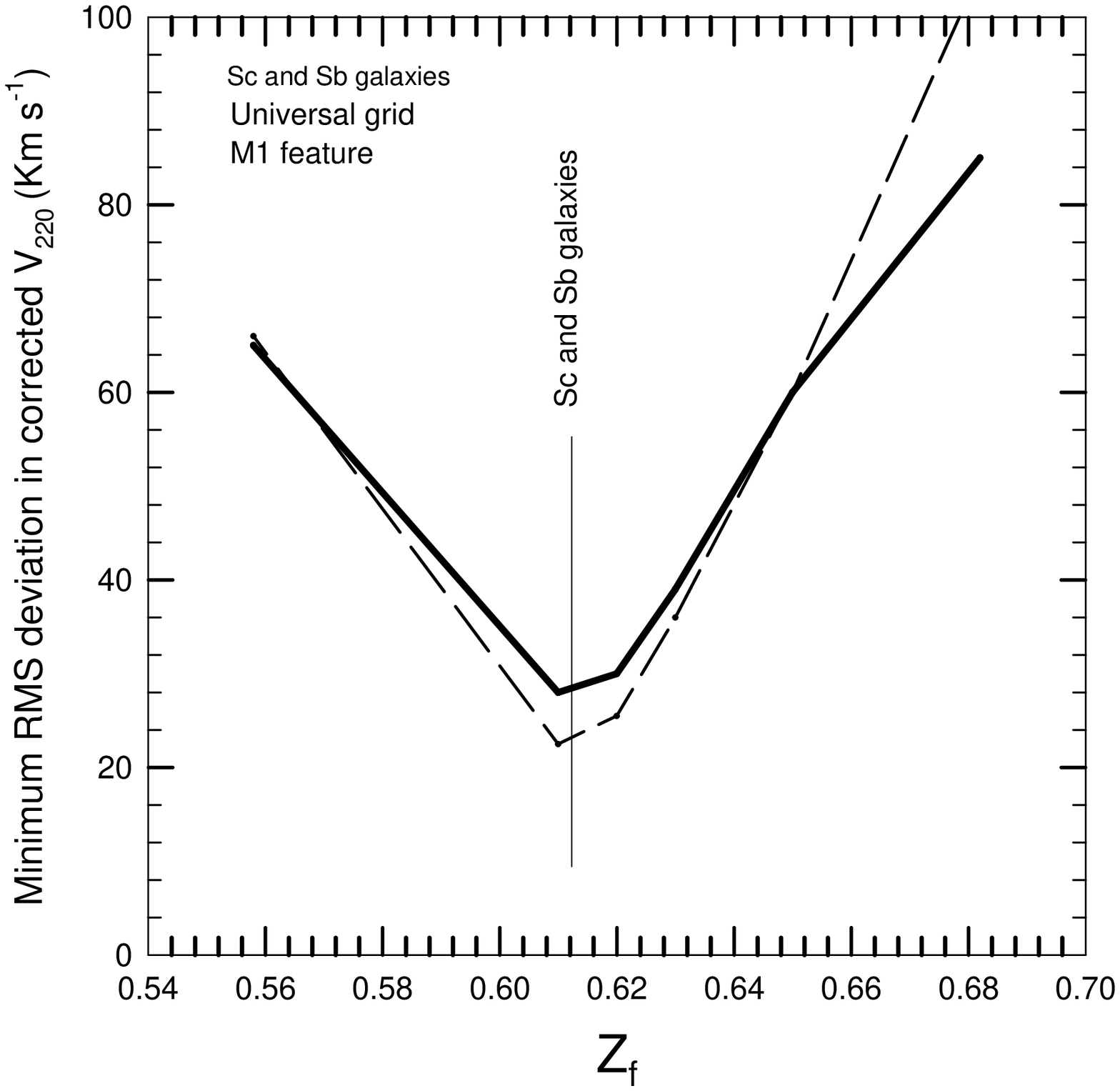}
\caption{\scriptsize{Plot of the minimum RMS deviation in corrected V$_{\rm 220}$ velocities as a function of z$_{f}$ for Sc and Sb galaxies in Tables 6 and 7 of \citet{rus02a}. Solid line is for Sc galaxies. \label{fig18}}}
\end{figure}

\begin{figure}
\hspace{-1.0cm}
\vspace{-1.0cm}
\epsscale{1.0}
\plotone{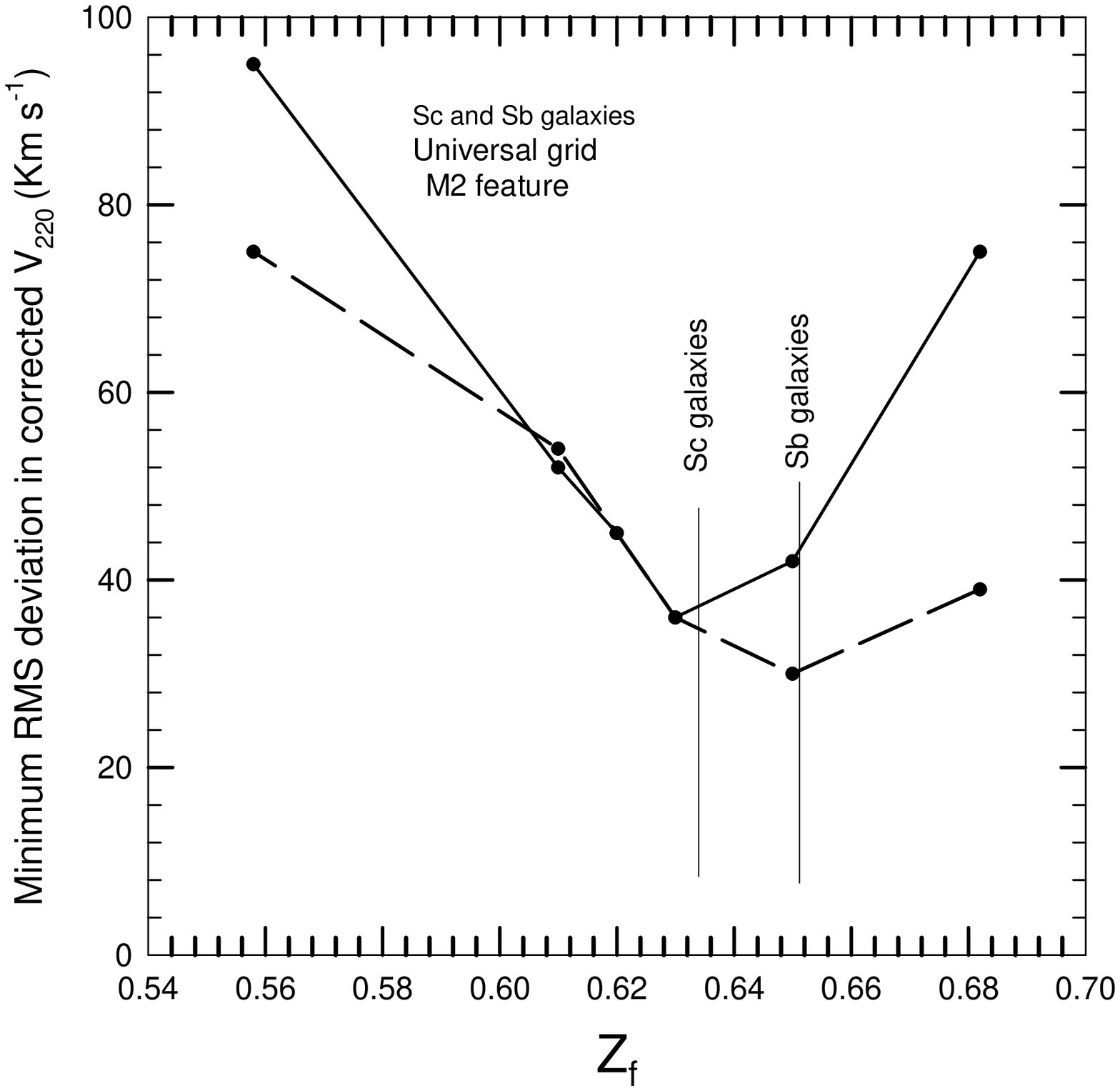}
\caption{\scriptsize{Plot of the minimum RMS deviation in corrected V$_{\rm 220}$ velocities as a function of z$_{f}$  for Sc and Sb galaxies in Table 6 and 7 of \citet{rus02a}. Solid line is for Sc galaxies. \label{fig19}}}
\end{figure}

\section{Intrinsic Redshift States of Spiral Galaxies}
 
The intrinsic redshift states predicted from earlier results \citep{tif96,tif97,bel02d,bel03a}, are listed here in Table 4. The levels assigned to the sources in the two spiral galaxy groups studied here are listed in Tables 5, and 6. Because the distance uncertainties in the Sb galaxies are known to be larger, some of the levels identified in Table 6 may have been misidentified. The states found for the FP clusters are listed in Table 7. Of the 66 sources/clusters fitted, at least 25 appear to be in the $N$ = 1 state (possibly as many as 32) and 18 are in the $N$ = 2 state. The spiral galaxies thus appear to contain a high percentage of sources in the $N$ = 1 state. This state corresponds to the T = 0 family of periods reported by \citet{tif96,tif97} and this result thus confirms the result found by Tifft that the T = 0 family of periods was the most common.   

\section{Distribution of Intrinsic Redshifts}

If the scatter in the standard redshift versus distance Hubble plot is due either to uncertainties in distance or the presence of peculiar velocities, it would be expected to show a Gaussian distribution about the Hubble line. We have argued previously, and have attempted to show here, that the scatter in the Hubble plot is produced, for the most part, by the presence of quantized intrinsic redshifts superimposed on the Hubble flow. In this scenario a Gaussian distribution about the Hubble line would not be expected, since the number of discrete redshift components per unit redshift decreases with increasing intrinsic redshift. Instead, a source distribution that bunches up at the low intrinsic redshift end (high $m$ numbers) would be more likely. In fact, this characteristic is visible in the data, and in Fig 21 the source distribution is shown as a function of excess redshift for the Sc, Sb, and FP galaxies. The filled dots indicate the location of the intrinsic redshifts in the $N$ = 1 state. Since these get very close together at high $m$-values, values $>$ [1,8] are included in one final level (the $m$ = [1,10] level = 18 km s$^{-1}$).  The excess is defined here as that portion of the redshift above the Hubble flow component, where the latter is the best-fit Hubble value we obtained for each group of galaxies. The choice of H$_{\rm o}$ does not significantly affect the shape of the distribution as long as reasonable values are assumed. In Fig 21 it is clear that the excess redshifts are heavily weighted towards the low intrinsic redshift side, and the distribution is clearly non-Gaussian. This result itself is a strong argument in favor of the existence of intrinsic redshifts.

\section{Intrinsic Redshifts in Type Ia Supernova Galaxies}

We have also examined the 36 SnIa galaxies studied by \citet{fre01}, and the results will be presented in \citet{bel03b}. However, Fig 22 has been copied from that paper and shows a plot of the RMS deviation in V$_{\rm CMB}$ vs H$_{\rm o}$ for the SnIa galaxies listed in Table 6 of \citet{fre01}. A best-fit feature is visible at H$_{\rm o}$ = 58 km s$^{-1}$ Mpc$^{-1}$. Thus the 91 spiral and SnIa galaxies all show approximately the same Hubble value of H$_{\rm o} = 58\pm2$.

\begin{figure}
\hspace{-1.0cm}
\vspace{-1.0cm}
\epsscale{1.0}
\plotone{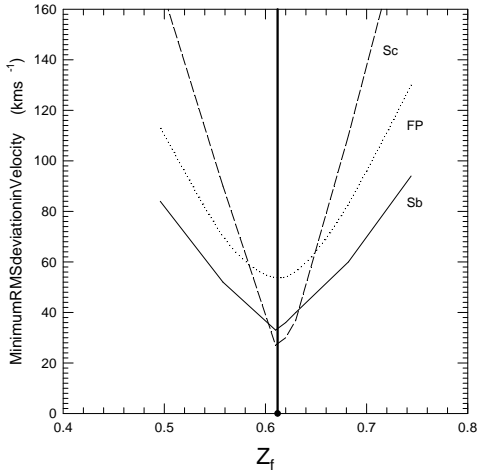}
\caption{\scriptsize{Plot of RMS deviation in velocity about the relevant intrinsic redshift grid lines as a function of the z$_{f}$ for the Sc, Sb, and FP groups. \label{fig20}}}
\end{figure}

\begin{figure}
\hspace{-0.5cm}
\vspace{-1.5cm}
\epsscale{1.0}
\plotone{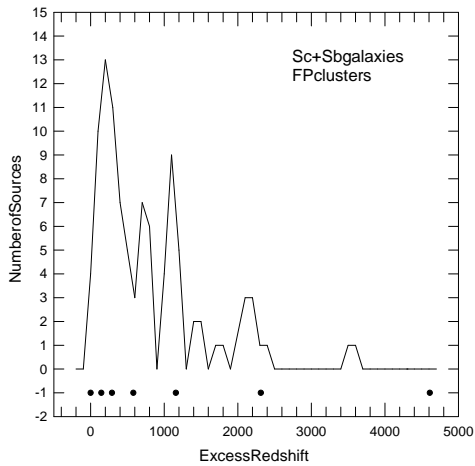}
\caption{\scriptsize{Plot of the relative density of excess redshifts for Sc, Sb, and FP galaxies/groups. The filled circles indicate the positions of the $N$ = 1 intrinsic redshifts and demonstrate how both the transitions and the excess "velocities" bunch up at high $m$-values. This shows clearly that the excess "velocity" distribution is not Gaussian as would be expected if they were random peculiar velocities. \label{fig21}}}
\end{figure}

\begin{figure}
\hspace{-1.0cm}
\vspace{-1.5cm}
\epsscale{1.0}
\plotone{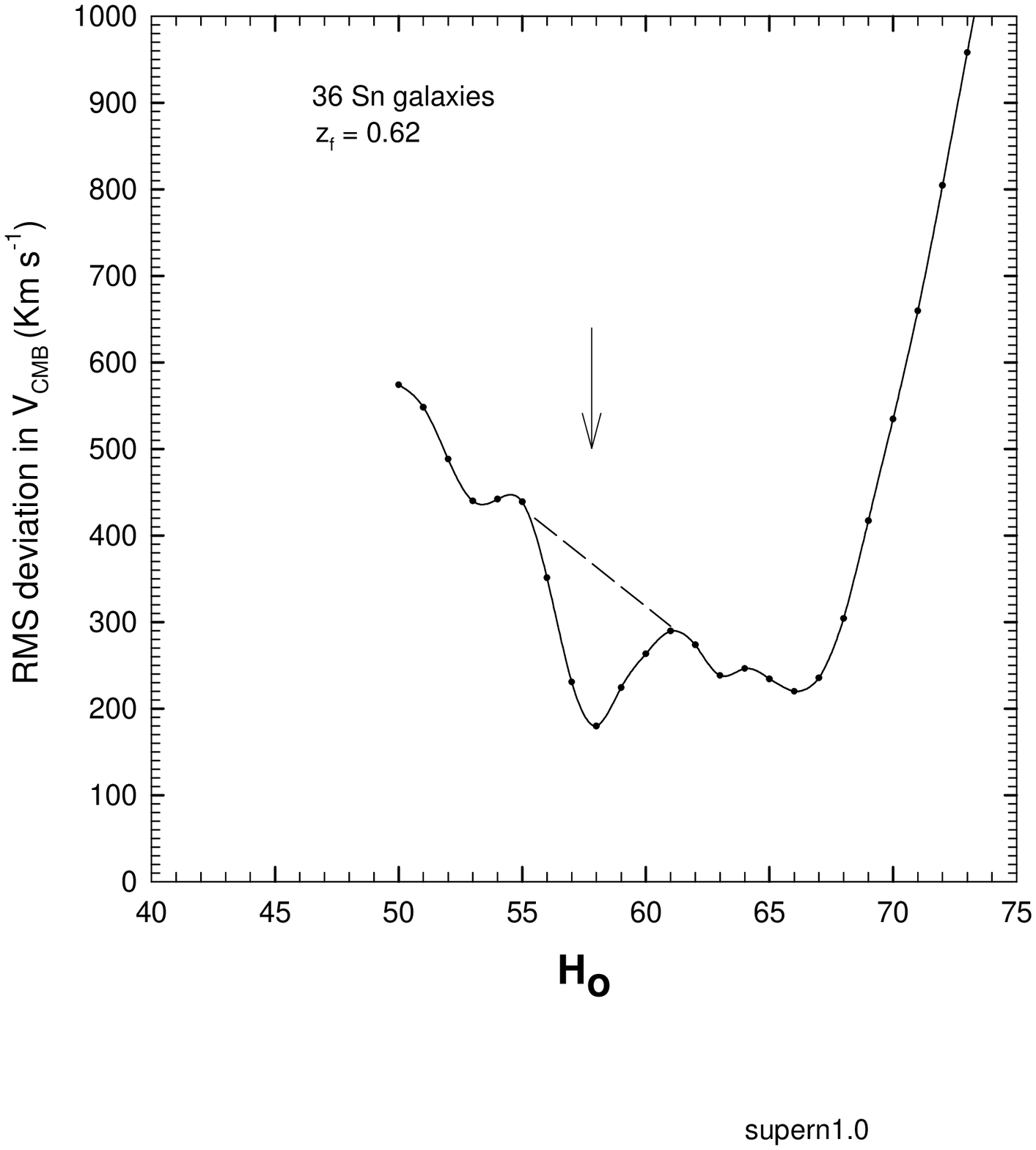}
\caption{\scriptsize{Plot of RMS deviation in V$_{\rm CMB}$ vs H$_{\rm o}$ for SnIa galaxies and z$_{f}$ = 0.62. Data are from Table 6 of \citet{fre01}. \label{fig22}}}
\end{figure}

\begin{figure}
\hspace{-1.0cm}
\vspace{-1.5cm}
\epsscale{1.1}
\plotone{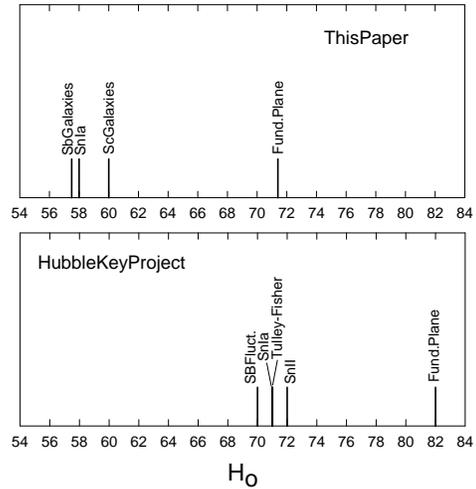}
\caption{\scriptsize{Comparison of H$_{\rm o}$-values found here with those found by the Hubble Key Project for different galaxy groups. This figure shows clearly that the mean Hubble constant obtained here is 20 percent lower than that obtained by the Hubble Key Project. \label{fig23}}}
\end{figure}

In Fig 23 the H$_{\rm o}$-values obtained for all groups are compared to those obtained by the Hubble Key Project. It shows that in both cases the FP results are discrepant, and that our values are systematically 20 percent lower for all groups. If the redshifts of these galaxies contain an intrinsic component as argued here, we therefore conclude that the local Hubble constant is likely to be closer to H$_{\rm o}$ = 58 than 72. This value is in good agreement with values near 60 found for the intermediate epoch (z $\sim$ 0.5) using the Sunyaev-Zel'dovich effect \citep{jon01,mas01,ree02}. It is also closer to the current epoch value near 55 claimed by \citet{san96,san00}.
%We also found a similar decrease in H$_{\rm o}$ for the Fundamental Plane %clusters where the Hubble Key Project value of 82 was found by us to be 71.4 %\citep{bel03a}.
This lower value for H$_{\rm o}$ is expected when the two techniques are compared, since the Hubble Key Project assumed that the intrinsic redshifts were peculiar velocities, or distance uncertainties, and included them in the Hubble constant determinations. Thus, as concluded previously, if the intrinsic redshifts claimed here are not taken into account when determining the Hubble constant, a much higher, but erroneous, value for H$_{\rm o}$ will be obtained.

\section{Is the Great Attractor a Myth?}

The Great Attractor (GA) model of Lynden-Bell et al. (1988) 
%and Faber & Burnstein (1988)
suggests that there is a large, approximately spherical overdensity of radius $\sim40$ $h^{-1}$ Mpc influencing the peculiar motions of galaxies over a large volume of space and is responsible for a large share of the Local Group's motion of 600 km s$^{-1}$ with respect to the CMB. The center of this over-dense region lies at a distance of approximately 4200 km s$^{-1}$ in the direction $l$ = 309\arcdeg, $b$ = 18\arcdeg. 

\citet{mat92a} surveyed the peculiar velocities of 1355 spiral galaxies visible in the southern hemisphere using the Tulley-Fisher (TF) relation to determine distances. They reported an excess of redshifted velocities relative to the CMB for these sources, most of which lay in the hemisphere containing the GA. One explanation for a velocity excess of this type \citep[Fig 15]{mat92a} is that it is due to the gravitational attraction of the GA. However, if intrinsic redshifts are real as argued here, excess redshifts like those found by \citet{mat92a} can also be explained by intrinsic redshifts superimposed on the Hubble flow. If the excess redshifts are really due to a directed flow towards the GA, they should become excess $blueshifts$ in the anti-GA direction. If they are due to intrinsic redshifts superimposed on the Hubble flow, they will also appear as excess $redshifts$ in the anti-GA direction. Which of these two scenarios is correct can thus be determined simply by observing sources in both directions. 

In the present study, and in that of \citet{bel03a}, we have found that the 'excess' components appear as redshifts for sources lying in all directions. Thus, when peculiar motions due to local density perturbations can be modeled and removed, these galaxies do not show an excess of blueshifts regardless of direction. What then is the evidence for a Great Attractor? \citet{mat92a} found clear evidence for an excess of redshifted 'peculiar velocities' relative to the Hubble flow in the Southern hemisphere which contains the Great Attractor. Although there have been independent studies of Northern Hemisphere galaxies, in making such comparisons it is desirable to use data that have been obtained in the same survey and analyzed by the same observers. Unfortunately the anti-GA direction was not covered in the Mathewson et al. survey. However, several sources, listed here in Table 3, were measured in the region near $l$ = 120\arcdeg, $b$ = -40\arcdeg (see Fig 1 of \citet{mat92a}). These sources lie only 10\arcdeg to 30\arcdeg from the anti-GA direction and are assumed here to be close enough to allow us to differentiate between a directed flow and intrinsic redshifts.

In Fig 24 these sources are shown on a Hubble plot. It is clear from this plot that the 'peculiar velocities' are not blueshifted as would be expected in the GA model, but redshifted as would be expected if they were intrinsic redshifts. We therefore suggest, barring the slight possibility that these objects just happen to be located near some other local overdensity, that when the same analysis process is used for both hemispheres, there is more evidence for intrinsic redshifts than for a directed flow towards a Great Attractor.

This is not the first time that the GA model has been questioned. \citet{mat92b} reported that they found no evidence for backside infall into the GA out to a distance of 60 $h^{-1}$ Mpc and concluded that although there appeared to be a bulk flow, there was no evidence that it was due to a GA. In the 4 different groups studied by us to date, when source velocities have been corrected for all known peculiar motions, any excess 'peculiar velocity' remaining has always been found to be positive (a redshift) for sources lying in all directions of the sky. This excess redshift, explained here by intrinsic redshifts, but previously assumed to be excess peculiar velocities, will have contaminated the data and may have prevented the determination of an accurate model. It is therefore suggested that if the excess redshifts are really intrinsic redshifts, the Great Attractor model may have come about because a), intrinsic redshifts were present but unaccounted for in the data and b), one of the main investigations involved sources in the Southern Hemisphere where intrinsic redshifts and a directed flow due to a 'great attractor' have similar characteristics. Although this is far from proven, we strongly suggest that the Great Attractor model needs to be re-examined.

\begin{figure}
\hspace{-1.0cm}
\vspace{-1.5cm}
\epsscale{1.0}
\plotone{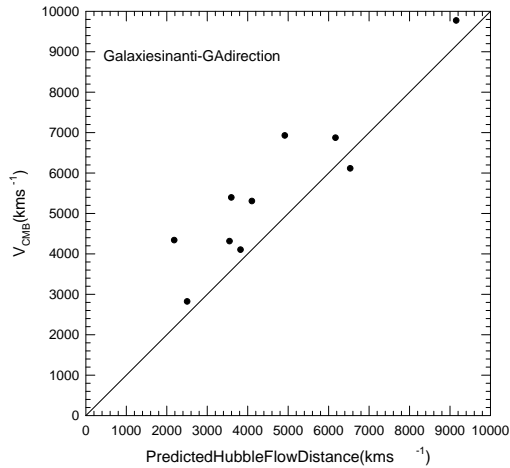}
\caption{\scriptsize{Plot of V$_{\rm CMB}$ vs  predicted Hubble flow distance  in km s$^{-1}$ for galaxies closest to the anti Great Attractor direction. Data are listed in Table 3 and are from \citet{mat92a}. \label{fig24}}}
\end{figure}

\begin{figure}
\hspace{-1.0cm}
\vspace{-1.0cm}
\epsscale{1.0}
\plotone{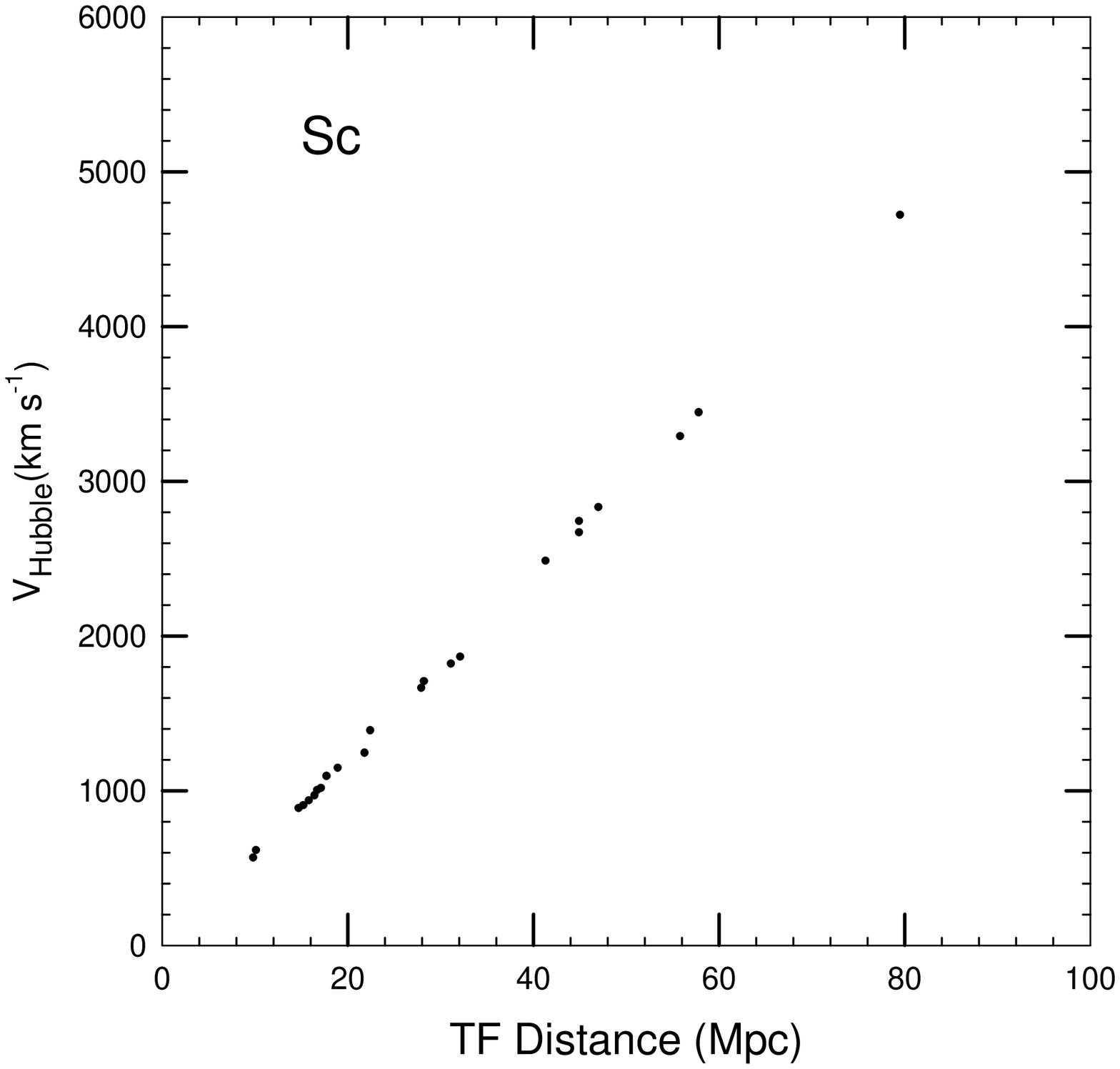}
\caption{\scriptsize{Plot of Hubble velocity vs TF distance for Sc galaxies after removal of intrinsic and V$_{\rm MW}$ velocities. \label{fig25}}}
\end{figure}

\begin{figure}
\hspace{-1.0cm}
\vspace{-1.0cm}
\epsscale{1.0}
\plotone{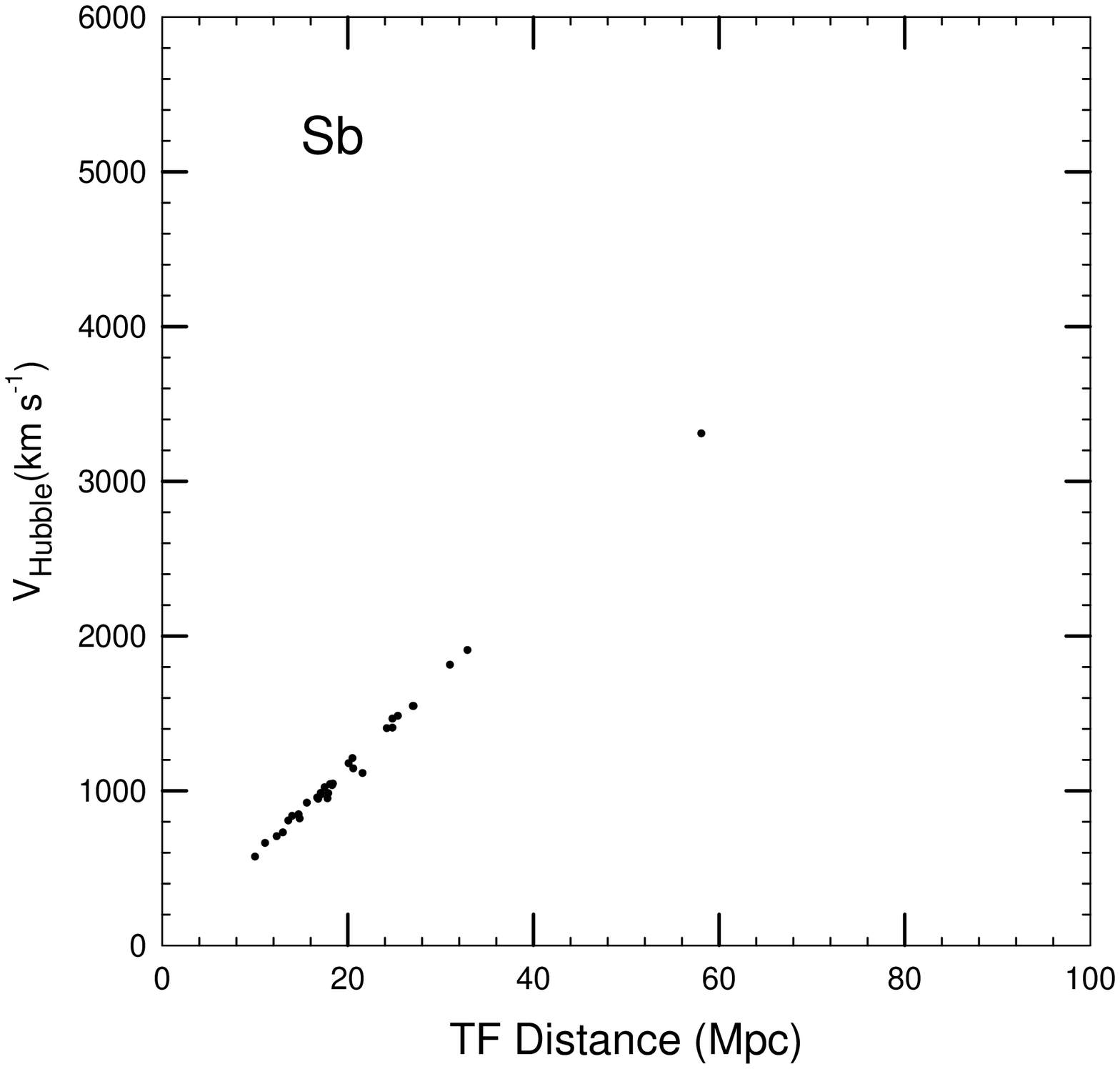}
\caption{\scriptsize{Plot of Hubble velocity vs TF distance for Sb galaxies after removal of intrinsic and V$_{\rm MW}$ velocities. \label{fig26}}}
\end{figure}

\begin{deluxetable}{ccccc}
\tabletypesize{\scriptsize}
\tablecaption{Summary of Parameters Found for Sc and Sb Galaxies.  \label{tbl-2}}
\tablewidth{0pt}
\tablehead{
\colhead{Group} & \colhead{H$_{\rm o}$ for M1} & \colhead{z$_{f}$ for M1}  & \colhead{H$_{\rm o}$ (mean of M1 and M2)} & \colhead{z$_{f}$ (mean of M1 and M2)}
}
\startdata
%Sn & 58.8  & 0.059 \\
Sc & 60.0 & 0.0613 & 58.75 & 0.0623 \\
Sb & 57.5 & 0.0613 & 56.25 & 0.0632 \\
%SBF & 67.5 & 0.063  \\
%FP & 71.4  & 0.061  \\
\enddata 
\end{deluxetable}

\begin{deluxetable}{ccccc}
\tabletypesize{\scriptsize}
\tablecaption{Galaxies Closest to Anti-GA Direction\tablenotemark{a} in \citet{mat92a} sample. \label{tbl-3}}
\tablewidth{0pt}
\tablehead{
\colhead{Source} & \colhead{$l$ (deg)} & \colhead{$b$ (deg)} & \colhead{Hubble Flow Distance (km s$^{-1}$)} & \colhead{V$_{\rm CMB}$} 
}
\startdata
UGC 14 & 109  & -38  & 6169  & 6873 \\
UGC 210 & 114  & -38  & 3822  & 4105 \\
UGC 321 & 117  & -39 & 3549  & 4315  \\
NGC 280 & 123 & -38  & 9154  & 9772 \\
UGC 541 & 123  & -41 & 4916  & 6927  \\
NGC 354 & 126  & -40 & 2183  & 4339  \\
NGC 697 & 140 & -38 & 2504  & 2822   \\
UGC 1934 & 150 & -34  & 6532  & 6116  \\
UGC 2020 & 151 & -34  & 4102  & 5308  \\
UGC 2079 & 152 & -33 & 3595  & 5394 \\
\enddata 
\tablenotetext{a}{anti-GA direction assumed is $l$ = 129\arcdeg, $b$ = -18\arcdeg}
\end{deluxetable}

\section{Discussion}

In Figs 25 and 26 the Hubble velocities from Table 5, col 5, and from Table 6, col 5, are plotted vs distance for the Sc and Sb galaxies respectively. Four Sb sources (NGC 1515, NGC 2683, NGC 4050, NGC 4548) have not been included in Fig 26 because they fall below the Hubble line. Even without these sources, the scatter is larger for the Sb galaxies, as anticipated above from the uncertainty in the distance modulus.

In our model the intrinsic redshifts are superimposed on top of the cosmological redshift due to the Hubble flow. It thus differs significantly from other models involving intrinsic redshifts where the entire redshift is assumed to be intrinsic \citep{kar71,kar77,nar80,tif02}. Thus if the results reported here are correct, all models based on the assumption that the entire redshift is intrinsic (and this includes 'tired light' models, etc.) would appear to be ruled out.

\section{Conclusions}

We have examined the Hubble plots of Sc, Sb and SnIa galaxies containing a total of 91 galaxies. We have found that their redshifts show evidence for quantized intrinsic redshift components superimposed on the Hubble flow that are related to those reported previously by Tifft and by us. We also find evidence that the distance uncertainties in the Sc group are smaller than those in the Sb group, and also that they are sufficiently small so that the detection of intrinsic redshift equivalent velocities at least as small as $\sim100$ km s$^{-1}$ is possible. This corresponds to a mean distance uncertainty of $\sim1.7$ Mpc. We find for the Sc, Sb, and SnIa groups, containing a total of 91 galaxies, that the local Hubble constant is H$_{\rm o}$ = $58\pm2$ km s$^{-1}$ Mpc$^{-1}$. This is 20 percent lower than the value (H$_{\rm o} = 72$) found by the Hubble Key Project, and is closer to the value near 60 km s$^{-1}$ Mpc$^{-1}$ reported for the intermediate Universe using the Sunyaev-Zel'dovich effect. The uncertainty in these numbers may still be as large as that quoted by the Hubble Key project. These groups also provide additional confirmations of the importance of the redshift increment z$_{f}$ = 0.62. They also offer a further confirmation of the quantized galaxy "velocities" reported by Tifft.

We thank Dr. D. McDiarmid for helpful comments when this manuscript was bring prepared.

\clearpage
\appendix

\begin{deluxetable}{ccccccccccccc}
\tabletypesize{\scriptsize}
\tablecaption{Redshifts and Velocities $N$ = 1 to 6. \label{tbl-4}}
\tablewidth{0pt}
\tablehead{
\colhead{$m$} & \colhead{z:$N$=1} & \colhead{vel:$N$=1} & \colhead{z:$N$=2} & \colhead{vel:$N$=2} & \colhead{z:$N$=3} & \colhead{vel:$N$=3} & \colhead{z:$N$=4} & \colhead{vel:$N$=4} & \colhead{z:$N$=5} & \colhead{vel:$N$=5} & \colhead{z:$N$=6} & \colhead{vel:$N$=6} 
}
\startdata
0 & 0.062  & 18012  & 0.31 & 79040 & 0.558 & 124852  & 1.178 & 195402 & 3.038 & 265145  & 3.658  & 273375 \\
1 & 0.031  & 9150    & 0.155 & 42904 & 0.279 & 72319  & 0.589  & 129693 & 1.519  & 218164  &  1.829  & 233195 \\
2 & 0.0155  & 4610.7 & 0.0775  & 22336  & 0.140 & 37058 & 0.294 & 75601 & .7595 & 153403  & .9145 & 171272 \\
3 & 0.00775 & 2314.3 & 0.0388  & 11392  & 0.0698 & 20197 & 0.147 & 40860 & .3797  & 93293  &.5472  & 123123 \\
4 & 0.00387 & 1157.9 & 0.0194 & 5752  & 0.0349 & 10280 & 0.0736 & 21254 & .1899 & 51605 & .2286 & 60862 \\
5 & 0.0019 & 580.13  & .009689 & 2890   & 0.0174 & 5171.0 & 0.0368 & 10829 & .0949 & 27106 & .1143 & 32319 \\
6 & 0.000969 & 290.36 & .00484  & 1449 & 0.00872 & 2602 & 0.0184 & 5465.4 & .0475  & 13902 & .0572 & 16658 \\
7 & 0.000484 & 145.15 & .00242 & 725.16  & 0.00436 & 1304 & 0.0092 & 2745.4 & .0237  & 7021 & .0286 & 8443 \\
8 & 0.000242 & 72.54  & .00121  & 362.8 & 0.00218 & 652.8 & 0.0046 & 1375.8 & .01187  & 3537 & .0143  & 4256 \\
9 & 0.000121 & 36.27  & .000605 & 181.3  & 0.00109  & 326.6 & 0.0023 & 688.7 & .00593 & 1772 & .00714 & 2133  \\
10 & 0.0000605 & 18.15 & .000303 & 90.7 & 0.000545 & 163.3 & 0.00115 & 344.5 & .002967  & 888 & .00357 & 1068 \\
11 & 0.0000303 & 9.054 & .000151 & 45.3 & 0.000272 & 81.53 & .000575& 172.3 & .001483 & 444 & .000893 & 534.8 \\
12 & 0.0000151 & 4.54  & .0000757 & 22.7 & 0.000136 & 40.77 & .000287 & 86.0 & .000742 & 222.4 & .000893 & 267.6  \\
13 &  --- & 2.27 & .0000378 & 11.3 & 0.0000681 & 20.42 & .0001437 & 43.1 & .00037 & 110.9 & .000447 & 133.8  \\
14 &  ---  & ---   & --- & ---  & 0.0000341 & 10.22 & .0000718 & 21.5 & .000189 & 56.6 & .000223 & 68.8\\
%15 & ---    & ---  & --- & --- & --- & --- & --- & --- & .0000945 & 28.3 & %.000112 & 33.57 \\
\enddata 
\tablenotetext{a}{velocities adjusted for relativistic effect}

\end{deluxetable}

\begin{deluxetable}{ccccc}
\tabletypesize{\scriptsize}
\tablecaption{Source Parameters for Sc Galaxies. \label{tbl-5}}
\tablewidth{0pt}
\tablehead{
\colhead{Cluster/Group} & \colhead{D (Mpc)} & \colhead{Corrected V$_{\rm 220}$ (km s$^{-1}$)\tablenotemark{a}} & \colhead{Transit.(Intrin. Vel.)(km s$^{-1}$)\tablenotemark{b}} & \colhead{V$_{\rm H}$ (km s$^{-1}$)\tablenotemark{c}}
}

\startdata
NGC309  & 32.1  & 5403 & z$_{\rm iG}$[5,8](3537)   & 1865.6  \\ 
NGC958  & 55.8  & 5423 & z$_{\rm iG}$[6,9](2133) & 3290.2  \\ 
NGC1232 & 18.9  & 1438 & z$_{\rm iG}$[1,6](290)    & 1147.9  \\ 
NGC1376 & 44.9  & 3901 & z$_{\rm iG}$[1,4](1158)    & 2742.8  \\ 
NGC2207 & 16.4  & 2418 & z$_{\rm iG}$[2,6](1449) & 969.1  \\ 
NGC2280 & 14.7  & 1611 & z$_{\rm iG}$[2,7](725)    & 886.0  \\ 
NGC2835 & 9.8  &  567   & z$_{\rm iG}$[1,10](18)    & 566.6  \\ 
NGC2942 & 57.8  & 4602 & z$_{\rm iG}$[1,4](1158)    & 3444.5  \\ 
NGC2955 & 79.5 & 7035  & z$_{\rm iG}$[1,3](2314) & 4720.9  \\ 
NGC2997 & 10.1 & 978  & z$_{\rm iG}$[2,8](363)    & 615.4   \\ 
NGC2998 & 44.9 & 4802  & z$_{\rm iG}$[6,9](2133) & 2668.7  \\
NGC3294 & 27.9  & 1953 & z$_{\rm iG}$[1,6](290)    & 1663.3  \\ 
NGC3464 & 47.0  & 3899 & z$_{\rm iG}$[6,10](1068)    & 2831.4  \\ 
NGC3614 & 28.2  & 2433 & z$_{\rm iG}$[2,7](725)    & 1707.8  \\ 
NGC3893 & 16.7  & 1115 & z$_{\rm iG}$[5,13](111)    & 1003.8  \\ 
NGC4254 & 17.1  & 2466 & z$_{\rm iG}$[2,6](1449) & 1016.9  \\ 
NGC4321 & 15.2  & 1630 & z$_{\rm iG}$[2,7](725)    & 905.0  \\ 
NGC4535 & 15.8  & 2005 & z$_{\rm iG}$[6,10](1068)    & 937.0  \\  
NGC4653 & 41.3  & 2631 & z$_{\rm iG}$[1,7](145)    & 2485.8  \\
NGC5161 & 21.8 & 2313 & z$_{\rm iG}$[6,10](1068)     & 1244.6  \\ 
NGC5230 & 31.1 & 6991 & z$_{\rm iG}$[3,5](5171)    & 1820.5  \\ 
NGC5364 & 17.7 & 1920 & z$_{\rm iG}$[2,7](725)     & 1094.8  \\
NGC6118 & 22.4 & 1834 & z$_{\rm iG}$[5,11](444)     & 1389.9  \\
\enddata
\tablenotetext{a}{total measured velocity including intrinsic component}
\tablenotetext{b}{intrinsic component obtained using z$_{f}$ = 0.62 and H$_{\rm o}$ = 60.}
\tablenotetext{c}{Hubble velocity after removal of intrinsic component and V$_{\rm MW}$}

\end{deluxetable}

\begin{deluxetable}{ccccc}
\tabletypesize{\scriptsize}
\tablecaption{Source Parameters for Sb Galaxies. \label{tbl-6}}
\tablewidth{0pt}
\tablehead{
\colhead{Cluster/Group} & \colhead{D (Mpc)} & \colhead{Corrected V$_{\rm 220}$ (km s$^{-1}$)\tablenotemark{a}} & \colhead{Transit.(intrin. Vel.)(km s$^{-1}$)\tablenotemark{b}} & \colhead{V$_{\rm H}$ (km s$^{-1}$)\tablenotemark{c}}
}

\startdata
NGC613  & 20.5  & 1179 & z$_{\rm iG}$[1,10](18)   & 1210 \\ 
NGC615  & 27.0  & 1553 & z$_{\rm iG}$[2,9](181)   & 1547  \\ 
NGC779  & 16.8  & 966  & z$_{\rm iG}$[1,6](290)   &  946  \\
NGC986  & 11.1  & 638 & z$_{\rm iG}$[6,10](1068)  &  661  \\ 
NGC1325 & 18.1  & 1041 & z$_{\rm iG}$[2,9](181)   & 1041  \\ 
NGC1964 & 18.3  & 1052 & z$_{\rm iG}$[6,12](268)   & 1034  \\ 
NGC2613 & 17.1  &  983 & z$_{\rm iG}$[1,5](580)   &  974   \\ 
NGC2712 & 27.1  & 1558 & z$_{\rm iG}$[6,12](268)   & 1547  \\ 
NGC2815 & 32.9  & 1892 & z$_{\rm iG}$[2,8](363)   & 1908  \\ 
NGC2985 & 20.6  & 1185 & z$_{\rm iG}$[6,11](535)   & 1143  \\
NGC3223 & 21.6  & 1242 & z$_{\rm iG}$[5,9](1772)  & 1113  \\ 
NGC3504 & 13.6  &  782 & z$_{\rm iG}$[1,4](1158)  &  807  \\ 
NGC3673 & 24.8  & 1426 & z$_{\rm iG}$[2,8](363)   & 1407  \\ 
NGC3675 & 16.7  &  960 & z$_{\rm iG}$[6,13](134)   & 954  \\ 
NGC3705 & 17.1  & 983  & z$_{\rm iG}$[1,10](18)   &  985  \\ 
NGC3717 & 17.9  & 1029 & z$_{\rm iG}$[5,10](888)   &  984  \\ 
NGC4679 & 31.0  & 1783 & z$_{\rm iG}$[2,5](2890)   & 1814 \\ 
NGC5054 & 17.5  & 1006 & z$_{\rm iG}$[2,7](725)    & 1022 \\
NGC5740 & 25.4  & 1461 & z$_{\rm iG}$[1,6](290)    & 1482 \\
NGC5850 & 17.8  & 1024 & z$_{\rm iG}$[5,9](1772)   &  950  \\ 
NGC5879 & 13.0  & 748  & z$_{\rm iG}$[2,7](725)    &  730 \\
NGC7177 & 20.1  & 1156 & z$_{\rm iG}$[2,8](363)    & 1176 \\ 
NGC7205 & 15.6  & 897  & z$_{\rm iG}$[5,11](444)    &  921 \\ 
NGC7217 & 18.4  & 1058 & z$_{\rm iG}$[2,9](181)    & 1045 \\ 
NGC7552 & 14.8  & 851  & z$_{\rm iG}$[3,8](653)    & 819 \\ 
NGC7723 & 24.2  & 1392 & z$_{\rm iG}$[5,11](444)    & 1403 \\ 
NGC7782 & 58.1  & 3341 & z$_{\rm iG}$[6,9](2133)   & 3308 \\ 
IC4351  & 24.8  & 1426 & z$_{\rm iG}$[1,4](1158)   & 1465 \\  
NGC3351 & 10.0  & 575  & z$_{\rm iG}$[6,13](134)    & 574 \\ 
NGC4725 & 12.36 & 711  & z$_{\rm iG}$[1,4](1158)   & 705 \\ 
NGC7331 & 14.72 & 845   & z$_{\rm iG}$[5,12](222)    & 847 \\
NGC2841 & 14.0  & 805  & z$_{\rm iG}$[1,10](18)    & 835 \\

\enddata
\tablenotetext{a}{V$_{\rm 220}$ corrected for V$_{\rm MW}$}
\tablenotetext{b}{intrinsic component obtained using z$_{f}$ = 0.62 and H$_{\rm o}$ =  57.5.}
\tablenotetext{c}{Hubble velocity after removal of intrinsic component}

\end{deluxetable}

\begin{deluxetable}{ccccc}
\tabletypesize{\scriptsize}
\tablecaption{Source Parameters for Fundamental Plane Clusters. \label{tbl-7}}
\tablewidth{0pt}
\tablehead{
\colhead{Cluster/Group} & \colhead{D (Mpc)} & \colhead{V$_{\rm CMB}$ (km s$^{-1}$)\tablenotemark{a}} & \colhead{Transit.(Disc.Vel.)(km s$^{-1}$)\tablenotemark{b}} & \colhead{V$_{\rm H}$ (km s$^{-1}$)\tablenotemark{c}}
}

\startdata
Dorado     & 13.8  & 1131 & z$_{\rm iG}$[1,7](145)  & 986  \\ 
Grm 15     & 47.4  & 4530 & z$_{\rm iG}$[1,4](1158) & 3372  \\ 
Hydra      & 49.1  & 4061 & z$_{\rm iG}$[1,5](580)  & 3481  \\ 
Abell S753 & 49.7  & 4351 & z$_{\rm iG}$[2,6](725) & 3626  \\ 
Abell 3574 & 51.6  & 4749 & z$_{\rm iG}$[1,4](1158) & 3591  \\ 
Abell 194  & 55.9  & 5100 & z$_{\rm iG}$[1,4](1158) & 3942  \\ 
Abell S639 & 59.6  & 6533 & z$_{\rm iG}$[1,3](2314) &  4219  \\ 
Coma       & 85.8  & 7143 & z$_{\rm iG}$[1,4](1158) &  5985  \\ 
Abell 539  & 102.0 & 8792 & z$_{\rm iG}$[2,7](1449) & 7343  \\ 
DC 2345-28 & 102.1 & 8500 & z$_{\rm iG}$[1,4](1158) &  7342  \\ 
Abell 3381 & 129.8 & 11536 & z$_{\rm iG}$[1,3](2314) &  9222  \\

\enddata
\tablenotetext{a}{Total measured velocity including intrinsic component}
\tablenotetext{b}{intrinsic component obtained using z$_{f}$ = 0.62.}
\tablenotetext{c}{Hubble velocity after removal of discrete component}

\end{deluxetable}

\end{document}